\documentclass[twocolumn,showpacs,preprintnumbers,pre,amsmath,amssymb]{revtex4}

\usepackage{graphicx}
\usepackage{dcolumn}
\usepackage{bm}
\usepackage{times}
\usepackage{float}
\usepackage{flafter}
\usepackage{epsfig}%

\newcommand{\gammadottot}{\dot{\gamma}_{\text{tot}}}
\newcommand{\gammadot}{\dot{\gamma}}

\newcommand{\gammadotco}{\dot{\gamma}_{\text{co}}}
\newcommand{\etaapp}{\eta_{\text {app}}}
\newcommand{\sigmadot}{\dot{\sigma}}

\newcommand{\rc}{r_{\text{c}}}
\newcommand{\sigmay}{\sigma_{\text y}}
\newcommand{\tauy}{\tau^{\text {y}}_{\text {dev}}}
\newcommand{\taurelax}{\tau_{\text{relax}}}

\newcommand{\sigmapeak}{\sigma^{\text{peak}}}
\newcommand{\sigmayt}{\sigma^{\text{yt}}}

\newcommand{\Tg}{T_{\text g}}
\newcommand{\Tc}{T_{\text c}}

\newcommand{\kh}{k_{\text h}}
\newcommand{\Uwall}{U_{\text wall}}
\newcommand{\Uthermal}{U^{\text{thermal}}}
\newcommand{\Umax}{U_{\text {max}}}

\newcommand{\tauco}{\tau_{\text co}}
\newcommand{\rhoA}{\rho_{\text A}}
\newcommand{\rhoB}{\rho_{\text B}}
\newcommand{\epot}{e_{\text {pot}}}

\newcommand{\myeq}{\!=\!}
\newcommand{\myapprox}{\!\approx\!}

\newcommand{\tw}{t_{\text {w}}}
\newcommand{\kB}{k_{\text {B}}}
\newcommand{\Xcm}{X_{\text {cm}}}
\newcommand{\Vcmmin}{V^{\text{min}}_{\text {cm}}}
\newcommand{\Vcm}{V_{\text {cm}}}
\newcommand{\Vcmel}{V^{\text{el}}_{\text {cm}}}

\newcommand{\snapshot}[2]{
\setlength{\unitlength}{1mm}
\begin{picture}(85, 45)(-#1, -#2)
\put(2,5){\epsfig{file=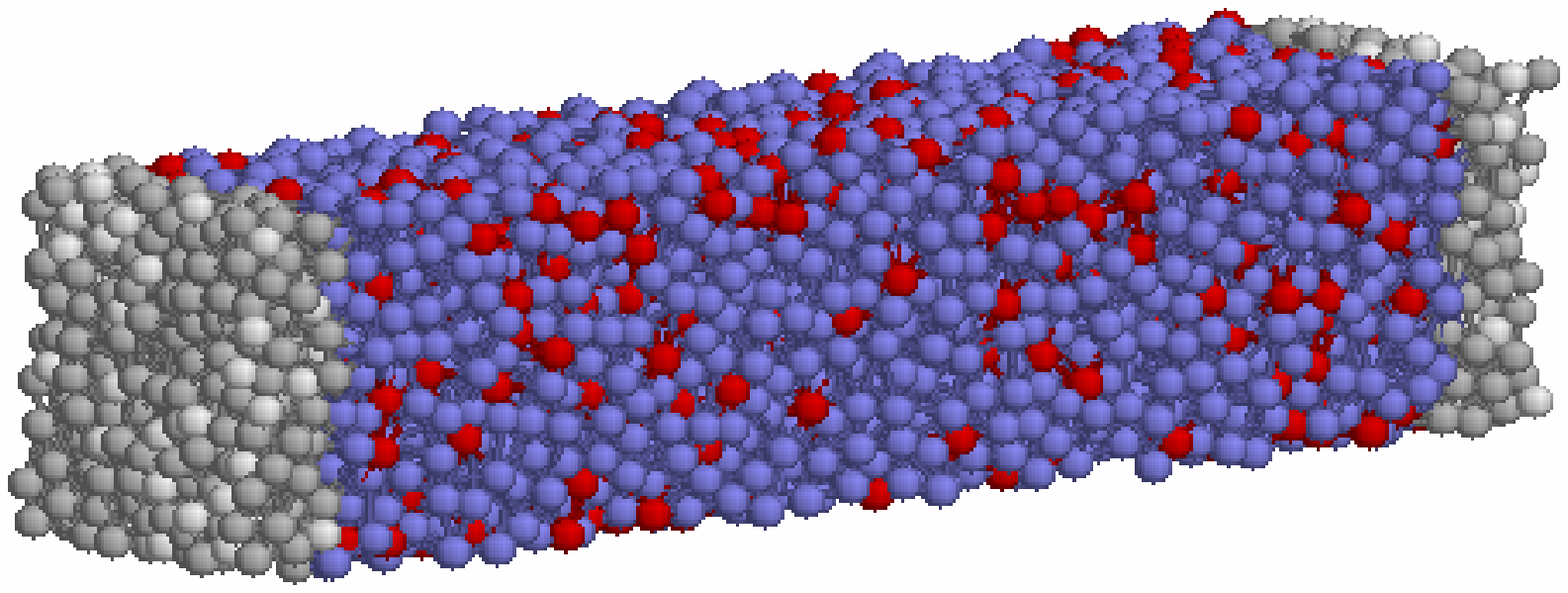, height=40mm, width=85mm,clip=, angle=0, silent=}}
\thicklines
\put(-3,17){$x$}
\put(0,0){\line(0,1){20}}

\put(17,-2){$y$}
\put(0,0){\line(5,-1){20}}

\put(30,3){$z$}
\put(0,0){\line(5,1){30}}
\end{picture}
}
\begin{document}
\title{A study of the static yield stress in a binary Lennard-Jones glass}
\author{F. Varnik$^{(1)}$, L. Bocquet$^{(2)}$, J.-L. Barrat$^{(2)}$}
\address{(1) CECAM, ENS-Lyon, 46 All\'ee d'Italie, 69007 Lyon, France\\
(2) Laboratoire de Physique de la Mati\`ere Condens\'e et
Nanostructures,
Universit\'e Lyon I and CNRS, 69622 Villeurbanne Cedex, France\\
}
\date{\today}

\begin{abstract}
The stress-strain relations and the yield behavior of a model
glass (a 80:20 binary Lennard-Jones mixtures~\cite{Kob-Andersen})
is studied by means of molecular dynamics simulations. In a
previous paper~\cite{VBBB} it was shown that, at temperatures
\emph{below}  the glass transition temperature, $\Tg$, the model
exhibits shear banding under imposed shear. It was also suggested
that this behavior is closely related to the existence of a
(static) yield stress (under applied stress, the system does not
flow until the stress $\sigma$ exceeds a threshold value
$\sigmay$). A thorough analysis of the static yield stress is
presented via simulations under imposed stress. Furthermore, using
steady shear simulations, the effect of physical aging, shear rate
and temperature on the stress-strain relation is investigated. In
particular, we find that the stress at the yield point (the
``peak''-value of the stress-strain curve) exhibits a logarithmic
dependence both on the imposed shear rate and on the ``age'' of
the system in qualitative agreement with experiments on amorphous
polymers~\cite{Govaert,vanAken} and on metallic
glasses~\cite{HoHuu,Johnson}. In addition to the very observation
of the yield stress which is an important feature seen in
experiments on complex systems like pastes, dense colloidal
suspensions~\cite{DaCruz} and foams~\cite{Debregeas::PRL87},
further links between our model and soft glassy materials are
found. An example are hysteresis loops in the system response to a
varying imposed stress. Finally, we measure the static yield
stress for our model and study its dependence on temperature. We
find that for temperatures far below the mode coupling critical
temperature of the model ($\Tc\myeq 0.435$), $\sigmay$ decreases
slowly upon heating followed by a stronger decrease as $\Tc$ is
approached. We discuss the reliability of results on  the static
yield stress and give a criterion for its validity in terms of the
time scales relevant to the problem.
\end{abstract}


\maketitle

\vskip2pc

\section {Introduction}
\label{section:introduction} Despite the large diversity of their
microstructures, the so called soft glassy
materials~\cite{Sollich} like pastes, dense colloidal suspensions,
granular systems and foams exhibit many common rheological
properties. Once in a glassy or ``jammed'' state, these systems do
not flow, if a small shear stress is applied on them. For stresses
slightly above a certain threshold value (the yield stress,
$\sigmay$), however, they no longer resist to the imposed stress
and a flow pattern is
formed~\cite{Larson,Coussot-Raynaud-et-al::PRL88::2002,Bonn2,DaCruz}.

\begin{figure}
\hspace*{-1mm}\epsfig{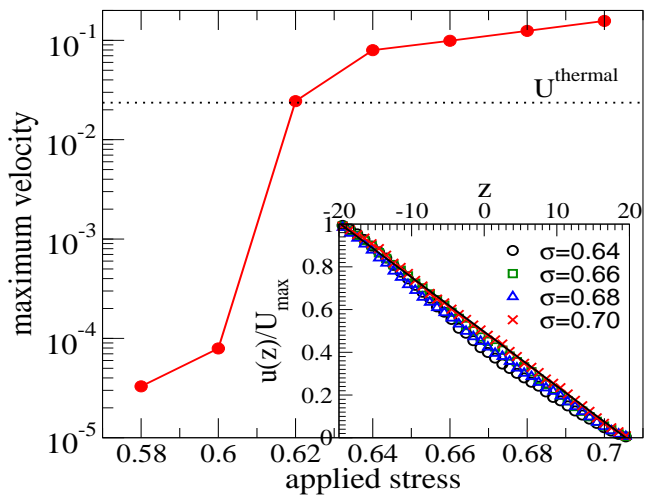}
\caption{\label{Vcm_Rcm_profile_dz1_sigma0.54-0.70} The maximum
velocity in the system, $\Umax$, measured in the layer of closest
approach to the left wall during simulations of a binary
Lennard-Jones glass [$T \myeq 0.2\; (<\Tc\myeq 0.435)$] at imposed
stress. The stress is increased by an amount of $d \sigma \myeq
0.02$ once in $4000$ LJ time units and $\Umax$ is measured between
two subsequent stress increments. The horizontal dotted line marks
the thermal velocity of the wall. Note the sharp increase in
$\Umax$ when changing the stress from $\sigma\myeq 0.6$ to $0.62$.
Inset: rescaled velocity profiles, $u(z) / \Umax $, measured
during the same simulations for stresses in the flow regime (for
which $\Umax \geq \Uthermal$). Obviously, once the flow sets in, a
linear velocity profile is formed across the system. }
\end{figure}

Let us illustrate this behavior using results of simulations to be
described in more detail in later sections.
Figure~\ref{Vcm_Rcm_profile_dz1_sigma0.54-0.70} shows $\Umax$, the
maximum velocity in the system measured close to the left wall
during simulations of a 80:20 binary Lennard-Jones (LJ)
mixture~\cite{Kob-Andersen} while applying a constant shear stress
to the left wall (see section~\ref{section:model} for more details
on the model). The applied stress is increased stepwise by an
amount of $d\sigma\myeq 0.02$  every $4000$ LJ time units and
$\Umax$ is measured between two increments of the stress (note
that, as seen from the inset of the same figure, this time is long
enough in order to also determine the velocity profile, $u(z)$,
accurately).

It is seen from Fig.~\ref{Vcm_Rcm_profile_dz1_sigma0.54-0.70}
that, for stresses $\sigma \leq 0.6$, $\Umax$ is hardly
distinguishable from zero. In particular, it is much smaller than
the thermal velocity of the wall, $\Uthermal \myeq \sqrt{T/M}
\myapprox 0.0236$ ($M\myeq 360$ is the  mass of the wall and
$T\myeq 0.2$ the temperature). Thus, at these stresses, the system
remains in the jammed state and resists to the drag force
transmitted to it by the left wall. However, as the stress is
further increased, a remarquable change in the system mobility is
observed. The system starts to flow and $\Umax$ increases by more
than two orders of magnitude.

An inspection of the corresponding velocity profiles illustrated
in the inset of Fig.~\ref{Vcm_Rcm_profile_dz1_sigma0.54-0.70}
reveals a further feature related to the yield stress, namely
that, once the applied stress exceeds the yield value, the whole
system fluidizes and the velocity profile is practically linear
(velocity profiles corresponding to $\sigma \leq 0.6$ fluctuate
around zero and are not shown in the inset).

On the other hand, in experiments upon imposed shear rate, shear
thinning is observed~\cite{Larson,Bonn}. The apparent viscosity,
defined as the average stress divided by the average overall shear
rate, $\etaapp \myeq \sigma/ \gammadottot$, decreases with
increasing $\gammadottot$  (in the case of a planar Couette-flow
with wall velocity and separation $\Uwall$ and $L_z$, for example,
$\gammadottot \myeq \Uwall/L_z$).  Furthermore, over some range of
shear rates, the system separates into regions with different
velocity gradients (shear
bands)~\cite{Bonn2,Coussot-Raynaud-et-al::PRL88::2002,Pignon::JRheo40::1996}.

Whereas the shear thinning is commonly attributed to the
acceleration of the intrinsic slow dynamics by the external flow
(the new time scale, $1/\gammadottot$, is much shorter than the
typical structural relaxation time of the
system)~\cite{Sollich,BB::PRE61::2000,BB2,Kurchan,Lacks}, the
origin of the shear bands still remains to be clarified. In some
cases, this shear-banding phenomenon can be understood in terms of
underlying structural changes in the fluid, analogous to a first
order phase transition. Examples are systems of rod like
particles, entangled polymers or surfactant micelles where the
constituents (rods, polymer or surfactant molecules) gradually
align with increasing shear rate thus leading to a coupling
between the local stress and the spatial variation of the velocity
gradient~\cite{olmsted,Dhont}. In the case of soft glassy
materials, however, no such changes are evident, and coexistence
appears between a completely steady region (zero shear rate)  and
a sheared, fluid
region~\cite{Chen,Pignon::JRheo40::1996,Losert::PRL85::2000,Debregeas::PRL87,Coussot-Raynaud-et-al::PRL88::2002}.

\begin{figure}
\hspace*{-4mm}\epsfig{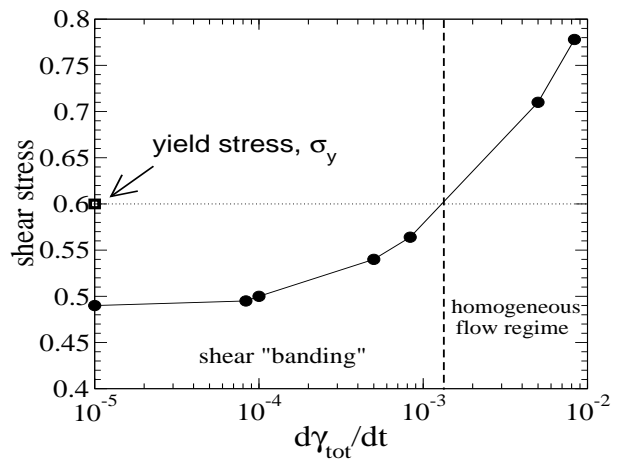} \caption{\label{fig:fig3} The shear
stress versus imposed shear rate, $\gammadottot$, under
homogeneous flow conditions at $T=0.2$. The square on the
horizontal axis marks the yield stress measured in imposed stress
simulations of a planar Couette cell [see the dramatic change in
$\Umax$ at $\sigma\myeq 0.6$ in
Fig.~\ref{Vcm_Rcm_profile_dz1_sigma0.54-0.70}, see also
Fig.~\ref{fig:yield_stress_versus_T}]. Under imposed shear, if the
corresponding steady state stress falls below the horizontal
dotted line, a heterogeneous flow can be expected, whereas in the
opposite case the flow will be homogeneous. The vertical dashed
line marks the shear rate on the boundary of these two flow
regimes. Note that the yield stress shown here is a \emph{lower}
bound for $\sigmay$ (see the solid line in
Fig.~\ref{fig:yield_stress_versus_T}) and thus is smaller than the
value used in~\cite{VBBB}. However, as a comparison with Fig.~3 of
Ref.~\cite{VBBB} shows, the estimated $\gammadottot$-range for
heterogeneous and homogeneous flow regimes is hardly changed by
this modification. }
\end{figure}

It was shown in a previous work~\cite{VBBB} that a model of 80:20
binary Lennard-Jones glass~\cite{Kob-Andersen} also exhibits the
shear banding phenomenon. Furthermore, a link was suggested
between the occurence of shear bands and the existence of a static
yield stress in the system. It was found that [see
Fig.~\ref{fig:fig3}] the yield stress is larger than the steady
state stress measured in a steady shear experiment in the limit of
the zero  shear rate, $\sigmay > \lim_{\gammadottot \to 0}
\sigma$. It was then suggested that, a shear-banding could be
expected for shear rates, for which
$\sigma(\gammadottot)<\sigmay$: as the flow is imposed externally
(by moving, say, the left wall) the formation of a flow pattern is
unavoidable. On the other hand, it follows from
$\sigma(\gammadottot) < \sigmay$ that, some regions in the system
are ``rigid enough'' to resist to the flow-induced stress whereas
other regions undergo irreversible rearrangements more
easily~\cite{footnote1}. Hence, whereas the details of the
``nucleation'' and growth of a heterogeneous flow pattern may
depend on the initial heterogeneity in the ``degree of
jamming''~\cite{CoussotJRheh46}, ``free volume''~\cite{Lemaitre}
or ``fluidity''~\cite{Derec,Picard} at the beginning of the shear motion,
its very origin lies in the possibility of resisting to the
shear-induced stress, i.e. in the existence of a static yield
stress.

Therefore, although it does not solve the problem of the selection
between the two bands, the existence of a static yield stress is
at least consistent with the coexistence of a jammed region and a
fluidized band: once the yield stress $\sigma_{\text y}$ is added
to the flow curve, the shear rate becomes multivalued in a range
of shear stresses, a situation encountered in several complex
fluids~\cite{olmsted}. This phenomenon should thus be generic for
many soft glassy materials.

In this paper we present an extensive study of the stress-strain
relations and yielding properties of the present model. The report
is organized as follows. After the introduction of the model in
the next section, results on the system response to an imposed
overall shear rate are presented, and the effects of physical
aging, shear rate and temperature on the stress-strain curves are
investigated. In section~\ref{section:imposedstress}, the response
of the system to imposed stress is studied. The measurement of the
static yield stress in the subject of
section~\ref{section:yieldstress}. A summary compiles our results.
\section {Model}
\label{section:model}
We performed molecular dynamics simulations of a generic glass
forming system, consisting of a 80:20 binary mixture of
Lennard-Jones particles (whose types we call A and B) at a total
density of $\rho\myeq \rhoA+\rhoB\myeq 1.2$. A and B particles
interact via a Lennard-Jones potential, $U_{\text{LJ}}(r)\myeq
4\epsilon_{\alpha\beta}[(\sigma_{\alpha\beta}/r)^{12}-(\sigma_{\alpha\beta}/r)^6]$,
with $\alpha,\beta\myeq {\text{A,B}}$. The parameters
$\epsilon_{\text{AA}}$, $\sigma_{\text{AA}}$ and $m_{\text{A}}$
define the units of energy, length and mass. The unit of time is
then given by $\tau\myeq
\sigma_{\text{AA}}\sqrt{m_{\text{A}}/\epsilon_{\text{AA}}}$.
Furthermore, we choose $\epsilon_{\text{AB}}\myeq
1.5\epsilon_{\text{AA}}$,  $\epsilon_{\text{BB}}\myeq
0.5\epsilon_{\text{AA}}$, $\sigma_{\text{AB}}\myeq
0.8\sigma_{\text{AA}}$, $\sigma_{\text{BB}}\myeq
0.88\sigma_{\text{AA}}$ and $m_{\text{B}}\myeq m_{\text{A}}$. The
potential was truncated at twice the minimum position of the LJ
potential, $\rc\myeq 2.245$. Note that the density is kept
constant at the value of $1.2$ for all simulations whose results
are reported here. This density is high enough so that the
pressure in the system is positive at all studied temperatures.
The present model system has been extensively  studied in previous
works~\cite{BB::PRE61::2000,barratkob,BB2,Kob-Andersen} and
exhibits, in the bulk state, a computer glass transition (in the
sense that the relaxation time becomes larger than typical
simulation times) at a temperature of $T_{\text c} \simeq
0.435$~\cite{Kob-Andersen}. Since our aim is to study the
interplay between the yield behavior and the possible flow
heterogeneities, we do not impose a constant velocity gradient
over the system as done in Ref.~\cite{BB2}, where a homogeneous
shear flow was imposed through the use of Lees-Edwards boundary
conditions. Rather, we confine the system between two solid walls,
which will be driven at constant velocity. By doing so, we mimic
an experimental shear cell, without imposing a uniform velocity
gradient.

We first equilibrate a large simulation box with periodic boundary
conditions in all directions, at $T\myeq 0.5$. The system is then
quenched to a temperature below $T_{\text c}$, where it falls out
of equilibrium, in the sense that structural relaxation times are
by orders of magnitude larger than the accessible simulation
times. On the time scale of computer simulation, the system is in
a glassy state, in which its properties slowly evolve with time
towards the (unreachable) equilibrium values (aging, see
Fig.~\ref{fig:msd_aging}). After a  time of $t\myeq 4.10^4$
[$2.10^6$ MD steps], we create  2 parallel solid boundaries by
freezing all the particles outside  two parallel $xy$-planes at
positions $z_{\text {wall}}\myeq \pm L_z/2$ ($L_z\myeq 40$) [see
Fig.~\ref{fig:snapshot}].  For each computer experiment, 10
independent samples (each containing 4800 fluid particles) are
prepared using this procedure. Note that the system is homogeneous
in the $xy$-plane ($L_x\myeq L_y\myeq 10$). We thus compute local
quantities like the velocity profile, the temperature profile,
etc. as an average over particles within thin layers parallel to
the wall.

\begin{figure}
\hspace*{-1mm}\epsfig{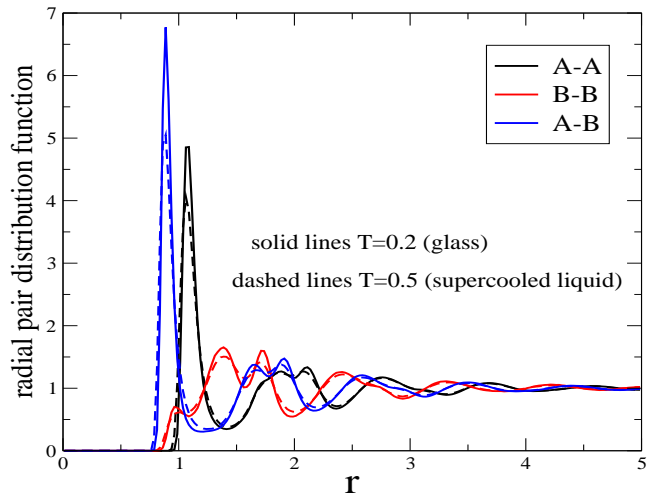}
\caption{\label{fig:pairdist}The radial pair distribution
functions, $g_{\alpha\beta}$ ($\alpha,\beta \in \{\text{A,B}\}$)
at two characteristic temperatures of $T\myeq 0.5$ (supercooled
state) and $T\myeq 0.2 $ (glassy state, note that the mode
coupling critical temperature of the system is $\Tc=0.435$).
Note that curves belonging to $T\myeq
0.5$ and $T\myeq 0.2$ are qualitatively similar (no further peaks
occur while cooling below $\Tc$). Thus, the system maintains its
liquid-like (amorphous) structure at temperatures  far below
$\Tc$.}
\end{figure}

\begin{figure}
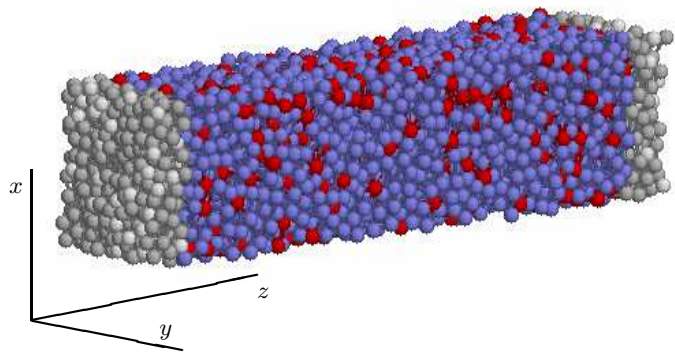

\snapshot{0}{5} \caption{\label{fig:snapshot} A snapshot of the
system at a total density of $\rho=\rhoA+\rhoB=1.2$ ($\rhoA=0.96$
and $\rhoB=0.24$). The walls (darker particles) are made of the
same types of particles as the fluid itself. They are
distinguished from the inner particles in that either they have no
thermal motion or they are coupled to equilibrium lattice sites by
harmonic springs thus preventing their diffusion.}
\end{figure}

\begin{figure}
\hspace*{-1mm}\epsfig{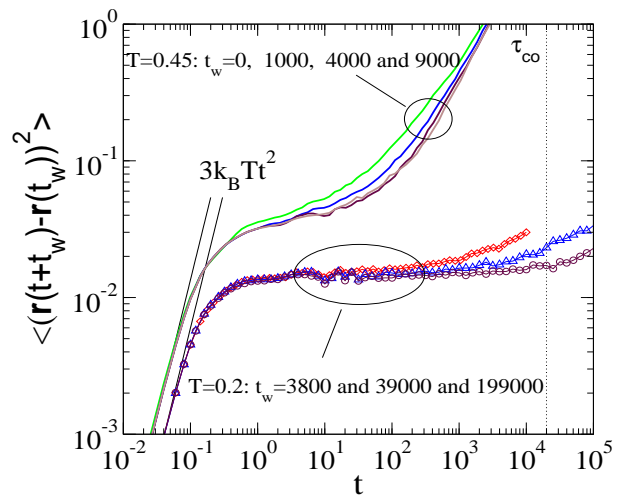} \caption{\label{fig:msd_aging}
The mean squared displacement (MSD) versus time at temperatures
$T\myeq 0.45$ (lines) and $T\myeq 0.2$ (symbols) for various
waiting times, $\tw$ (increasing from top to  bottom). $\tw \myeq
0$ corresponds to the time of temperature change from $T\myeq 0.5$
to the investigated temperature. While at $T\myeq 0.45$ the time
translation invariance (TTI) is reached after a few thousand LJ
time units, the data corresponding to $T\myeq 0.2$ indicate an
endless evolution towards slower dynamics. The larger $\tw$, the
wider the plateau and thus the longer the life time of the cage.
The straight lines are fits to the short time behavior of the
curves assuming free particle motion with thermal velocity. The
vertical dotted line marks $\tauco\myeq 2\times 10^{4}$. This time
is closely related to a change in the $\gammadottot$-dependence of
$\sigmapeak$ [see the discussion of Fig.~\ref{fig:sigmapeak}]. }
\end{figure}

The amorphous character of our model is clearly seen by an
analysis of the packing structure, i.e. the radial pair
distribution function. Figure~\ref{fig:pairdist} shows the various
kinds of radial pair distribution functions which can be defined
for a binary mixture: $g_{\alpha\beta}$ is the probability
(normalized to that of an ideal gas) of finding a particle of type
$\alpha$ at a distance $r$ of a $\beta$-particle ($\alpha,\beta
\in \{\text{A,B}\}$). In order to demonstrate that the system
keeps its amorphous structure at temperatures far below the glass
transition temperature of the model, we show the mentioned pair
distribution functions at two characteristic temperatures, one in
the supercooled state ($T \myeq 0.5 > \Tc \myeq 0.435$) and one at
$T\myeq 0.2$. As seen from Fig.~\ref{fig:pairdist}, the maxima of
$g_{\alpha\beta}$ are more pronounced at lower $T$. However, no
sign of crystallization or  long range positional order is
observed as the temperature is lowered through the glass
transition.

The mentioned insensitivity of the static structure to the glass
transition must be contrasted to the fact that, at temperatures
slightly above $\Tc$, the system can be equilibrated within the
time accessible to the simulation whereas this is no longer the
case for temperatures significantly below $\Tc$. At $T\myeq 0.5$,
for example, the time necessary for an equilibration of the system
is of order of a few hundred Lennard-Jones time  units (not
shown). For $T\myeq 0.45$, the equilibration time rises to a few
thousands whereas at $T\myeq 0.2$ the system is not equilibrated
even after $2\times 10^{5}$ LJ time units. At this temperature,
time translation invariance does not hold and the dynamical
quantities depend on \emph{two} times: the actual time, $t$, and
the waiting time $\tw$. Here, $\tw$ is the time elapsed after the
temperature quench (from $T\myeq 0.5$ to $T\myeq 0.2$) and the
beginning of the measurement.

This behavior is illustrated in Fig.~\ref{fig:msd_aging}, where
the mean square displacement (MSD) of a tagged particle is shown
at a temperature \emph{above} $\Tc$ ($T\myeq 0.45$) and at $T\myeq
0.2$ (far below $\Tc$) for various waiting times. The figure
nicely demonstrates the establishing of the time translation
invariance (TTI) at $T\myeq 0.45$. Here, $\tw\myeq 0$ corresponds
to a change of temperature from $T\myeq 0.5$ to $T\myeq 0.45$. As
expected from the fact that $T\myeq 0.45$ belongs to the
supercooled (liquid) state, with increasing waiting time, the MSD
converges towards the equilibrium curve reaching it after about
$4000$ Lennard-Jones time units. It is worth noting that the
waiting time at which the TTI is recovered roughly corresponds to
the time needed for the MSD to reach the size of a particle.

At $T\myeq 0.45$, the equilibrium curve for the MSD exhibits the
well-known two step relaxation characteristic of supercooled
liquid: for short times ($t\ll 1$), free particle motion with
thermal velocity is observed ($[\bm{r}(t+\tw)-\bm{r}(\tw)]^2 \myeq
(\bm{v}^{\text th}t)^2 \myeq 3\kB T t^2$). The free (ballistic)
motion ends up in a plateau thus indicating the (temporal) arrest
of the tagged particle in the cage formed by its neighbours.
Already after a few hundred LJ time units, the plateau is
gradually left and the MSD crosses over towards a linear
dependence on time (diffusive regime). This is indicative of
cooperative relaxation processes leading to the final release of
the tagged particle from the cage (cage relaxation).

At $T\myeq 0.2$, however, the situation is completely different.
Here,  TTI is not reached on the simulation time scale. Even after
a waiting time of $10^{5}$ LJ time units, the MSD continues
slowing down without reaching a steady state. The slowing down of
the dynamics with $\tw$ also has a direct consequence on the life
time of the cage. Figure~\ref{fig:msd_aging} shows that, as $\tw$
increases, so also does the width of the plateau. Hence, the time
necessary for the cage relaxation increases continuously with
$\tw$. However, in the case of $\tw\myeq 3.9\times 10^{4}$, one
can observe the very beginning of the cage relaxation around $t
\myapprox 2\times 10^{4}$. As will be discussed in
section~\ref{section:imposedshear}, this has an important
consequence for the shear rate dependence of $\sigmapeak$, the
stress at the maximum of stress-strain curves.

\section {Results at imposed shear rate}
\label{section:imposedshear}
An overall shear rate is imposed by moving in the $x$-direction,
say, the left wall ($z_{\text{wall}}\myeq -20$) with a constant
velocity of $U_{\text{wall}}$. This defines the total shear rate
$\gammadottot\myeq U_{\text{wall}}/L_z$. The motion of the wall is
realized in two different ways. One method used in our simulations
is to move all wall atoms with strictly the same velocity. In this
case, wall atoms do not have any thermal motion. As a consequence,
the only way to keep the system temperature constant, is to
thermostat the fluid atoms directly. A different kind of wall
motion is realized by coupling each wall atom to its equilibrium
lattice position via a harmonic spring~\cite{He-Robbins}. In this
case, the lattice sites are moved with a strictly constant
velocity while each wall atom is allowed to move according to the
forces acting upon it [the harmonic forces ensure that the wall
atoms follow the motion of the equilibrium lattice sites]. In such
a situation, we can thermostat the wall atoms while leaving the
fluid particles unperturbed. The temperature of the inner part of
the system is then a result of the heat exchange with the walls
(which now act as a heat bath). This method has the advantage of
leaving the fluid dynamics unperturbed by the thermostat.

The drawback of thermostating the system through the heat exchange
with the walls is that, depending on the shear rate and the
stiffness of the harmonic spring, measured by the spring constant
$\kh$, a temperature profile can develop across the system. Note
that the smaller the harmonic spring constant, the better the heat
exchange with the walls and thus the more efficient the system is
thermostated (the imposed shear rate having the opposite effect).
On the other hand, if $\kh$ is too small, the fluid particles may
penetrate the walls. We find that $\kh\myeq 25$ is a reasonable
choice for our model. However, even with this value of the
harmonic spring constant, we observe a temperature profile as the
shear rate exceeds $\gammadottot\myeq 10^{-4}$. For $\gammadottot
\myeq 10^{-3}$, for example, the maximum temperature in the fluid
is by about $3\%$ higher than the prescribed value.

In order to prevent such uncontrolled temperature increases, we
have therefore decided to apply direct thermostating to the inner
particles at all shear rates, independently of the possibility of
the heat exchange with the walls. For this purpose, we divide the
system into parallel layers of thickness $dz=0.25$ and rescale
(once every 10 integration steps) the $y$-component of the
particle velocities within the layer, so as to  impose the desired
temperature $T$. Such a local treatment is necessary to keep a
homogeneous temperature profile when flow profiles are
heterogeneous. To check for a possible influence of the
thermostat, we compared, for low shear rates ($\gammadottot\leq
10^{-4}$), these results with the output of a simulation where the
inner part of the system was unperturbed and the walls were
thermostatted instead. Both methods give identical results,
indicating that the system properties are  not affected by the
thermostat.

\begin{figure}
\hspace*{-1mm}\epsfig{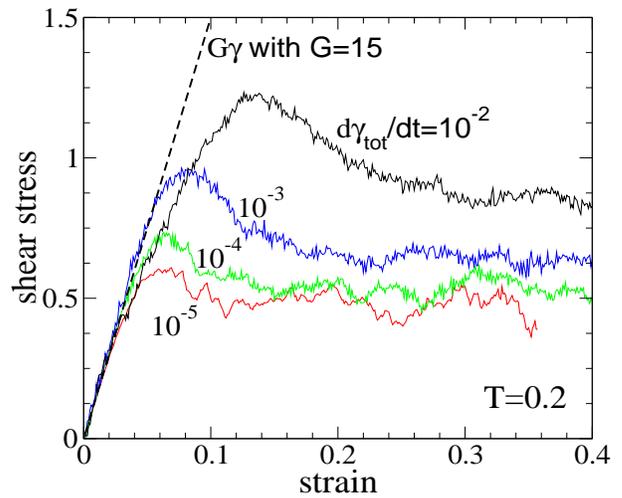}
\caption{\label{fig:stress-strain::gammadot_effect} Stress
response versus applied strain, $\gamma \myeq \gammadottot t$, for
the strain rates of $\gammadottot \myeq  10^{-5},\; 10^{-4},\;
10^{-3}$ and $10^{-2}$. Note that both the maximum and the steady
state values of the stress decrease with decreasing shear rate.
All stress-strain curves coincide with the straight line
$\sigma\myeq G \gamma$ ($G\myapprox 15$ is the elastic shear
modulus) at small deformations ($\gamma \le 2\%$). }
\end{figure}

However, for wall velocities close to $1$ or larger (corresponding
to overall shear rates of $\gammadottot \geq 2.5\times 10^{-2}$),
a non-uniform temperature profile develops across the system even
if the velocities are rescaled extremely
frequently~\cite{footnote2}. This can be rationalized as follows.
The heat created by the shear motion needs approximately $t_c\myeq
c/L_z$ to transverse the system ($c$ is the sound velocity). We
can estimate the sound velocity from a knowledge of the shear
modulus, $G$, and the density of the system, $c \myeq
\sqrt{G/\rho}$. At $T\myeq 0.2$ we find $G\myapprox 15$ (see
Fig.~\ref{fig:stress-strain::gammadot_effect}) thus obtaining
$c\myapprox 3.54$. A time of $t_c \approx 11.3$ is therefore
needed for a signal to transverse the whole system. Note that the
heat creation rate is given by $ dQ / dt \myeq \sigma
\gammadottot$ (neglecting inhomogeneities in the local shear
rate). An amount of energy equal to $\kB T$ is thus generated
within $t_Q \myeq \kB T / \dot{Q}$. The requirement $t_Q \geq t_c$
now means that the heat creation must be slow enough so that the
created energy can be dissipated in the whole system efficiently.
This gives $\gammadottot \leq \kB T / \sigma t_c$, which, after
setting $T\myeq 0.2$ and $\sigma \myapprox 0.6$, yields
$\gammadottot\leq 3\times 10^{-2}$.

Figure~\ref{fig:stress-strain::gammadot_effect} shows a typical
set of (transient) stress-strain curves at a temperature of
$T\myeq 0.2$ and for a waiting time of $\tw \myeq 4 \times 10^{4}$
LJ time units. The varying parameter is the overall shear rate
$\gammadottot\myeq \Uwall/L_z$ (the strain is simply computed as
$\gamma \myeq \gammadottot t$). First, an elastic regime is
observed at small shear deformations ($\gamma \leq 0.02$). The
stress then increases up to a maximum, $\sigmapeak$, before
decreasing towards the steady state stress at large deformations.
Therefore, this maximum is sometimes referred to as the yield
point~\cite{Utz} or dynamical yield stress~\cite{Khan}. In the
following, we will simply refer to this quantity as $\sigmapeak$,
since plastic (irreversible) deformation actually sets in before
the corresponding value of the strain is reached.  Moreover, as
will be seen below, $\sigmapeak$  depends on strain rate and
waiting time in a nontrivial way, so that it is difficult, in our
simulations,  to define a yield stress value from such dynamical
stress/strain curve.

 As commonly observed in experiments on
polymers~\cite{Govaert} and on metallic
glasses~\cite{vanAken,Johnson}, the stress overshoot $\sigmapeak$
decreases and is observed at smaller strains as the shear rate is
lowered [see also Fig.~\ref{fig:sigmapeak}]. Note also that all
curves in Fig.~\ref{fig:stress-strain::gammadot_effect} show the
same elastic response at small strains. As also shown in the
figure, a linear fit to $ \sigma\myeq G \gamma$ with a shear
modulus of $G\myapprox 15$ describes well the data at small
deformations.

\begin{figure}
\hspace*{-1mm}\epsfig{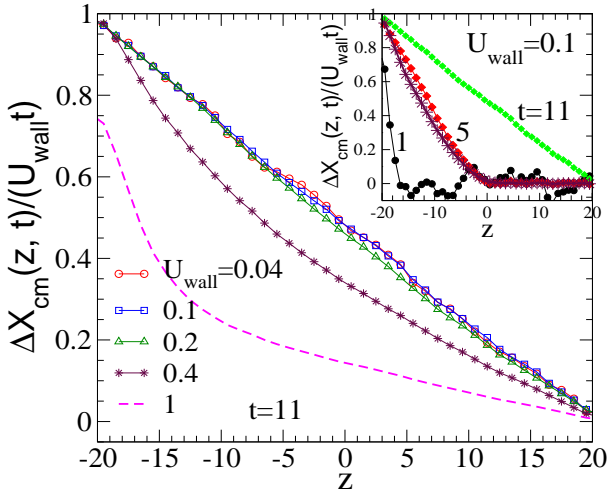}
\caption{\label{fig:short-time-Xcm_profile} Short time behavior of
the layer resolved displacements of the center of mass normalized
to the displacement of the moving wall, $(\Xcm(z; t)-\Xcm(z;
0))/(\Uwall t)$ ($z$ denotes the position of the middle of the
layer. The system is divided into layers of thickness $\Delta z
\myeq 1$ and $\Xcm$ is measured by averaging over the
$x$-coordinates of all particles within the specified layer). The
displacement field is shown at $t \myeq 11$ (note that the time
needed by the sound to travel across the system is given by
$t_c\myeq L_z /c \myapprox 11.3$) for various wall velocities as
indicated in the figure. For $\Uwall \leq 0.2$, a linear
deformation profile is observed, whereas at higher wall velocities
this is no longer the case. The inset shows, for a (low) wall
velocity of $\Uwall\myeq 0.1$, how the deformation field
propagates towards the immobile wall (placed at $z\myeq 20$). The
speed with which the boundary of the deformed region extends
towards the immobile wall is found to be indeed very close to the
estimated value of the sound velocity $c \myapprox 3.54$. The
stars in the inset correspond to $\Uwall \myeq 0.4$ at $t\myeq 5$
demonstrating that, at a time corresponding to a smaller strain,
the \emph{local} response  of the system is elastic [see also the
text for more discussion]. }
\end{figure}

\begin{figure}
\hspace*{-1mm}\epsfig{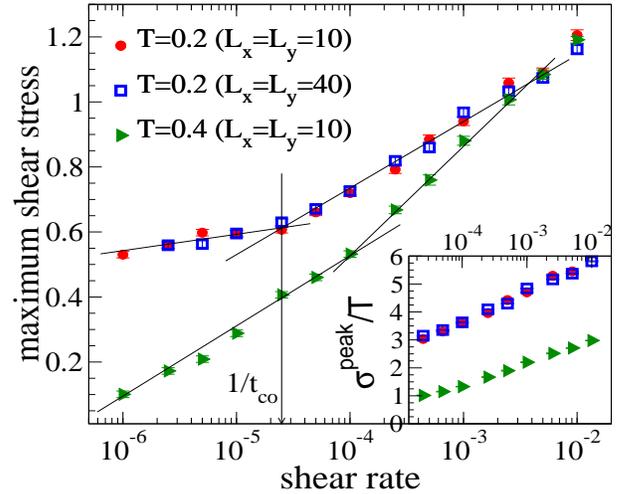} \caption{\label{fig:sigmapeak}
The maximum of the stress-strain curve, $\sigmapeak$, versus
strain rate. Data for $T\myeq 0.2$ correspond to two different
system sizes: $L_x\myeq L_y \myeq 10,\; L_z\myeq 40$ (averaged
over 10 independent runs) and  $L_x \myeq L_y \myeq L_z \myeq 40$
(a sole run). Apparently, results on $\sigmapeak$ do not depend
much on the system size. At the higher temperature ($T\myeq 0.4$),
only the smaller system size (again averaged over 10 independent
runs) is used. At approximately $\gammadottot \myeq 2.5 \times
10^{-5}$, the slope of $\sigmapeak$ changes significantly at
$T\myeq 0.2$. For $T\myeq 0.4$, a similar (albeit not so
strong) change in slope occurs at a higher shear rate. Solid lines
are guides for the eye. The inset shows the same data, where
$\sigmapeak$ divided by temperature is shown versus
$\gammadottot$. }
\end{figure}

In order to understand the rather strong deviation from linearity
at small strains in the case of $\gammadottot\myeq 10^{-2}$, we
recall that, once the (left) wall starts its motion, a time of
approximately $t_c \myeq 11.3$ must elapse before the deformation
field comprises the whole system. This is nicely borne out in the
inset of Fig.~\ref{fig:short-time-Xcm_profile} where, for a wall
velocity of $\Uwall\myeq 0.1$, ``snap shots'' of the layer
resolved displacement of center of mass (normalized to the
displacement of the wall) are shown for $t \myeq 1, \; 5$ and
$11$. Indeed, the boundary of the deformed region reaches the
immobile wall only after $t\myeq 11$ LJ time units. We have
verified this behavior for other wall velocities and have found
$t\myapprox 11$ in all cases. However, as shown in the main part
of Fig.~\ref{fig:short-time-Xcm_profile}, at higher wall
velocities, the deformation field is no longer linear at the time
it reaches the immobile wall. This can be rationalized as follows.
The total strain at $t\myeq t_c$ is given by $\gamma \myeq
\gammadottot t_c$ yielding $\gamma \myeq 11\%$ for $\gammadottot
\myeq 0.4/40 \myeq 10^{-2}$. Hence, the elastic regime is left
already before the whole system is affected by the motion of the
wall. Putting it the other way, one can estimate the time for
which a \emph{locally} elastic response can still be observed at a
given wall velocity: $t_{\text{el.resp.}}\myeq \gamma_{\text{el}}
/ \gammadottot$. Assuming an elastic response at a strain of a few
percent one obtains for $t_{\text{el.resp.}}$ a time of a few
Lennard-Jones units at $\gammadottot\myeq 10^{-2}$ [see the stars
in the inset of Fig.~\ref{fig:short-time-Xcm_profile}].

The dependence of $\sigmapeak$ on $\gammadottot$ is depicted in
Fig.~\ref{fig:sigmapeak} for temperatures of $T\myeq 0.2$ and
$T\myeq 0.4$. For the lower temperature, data are shown for two
system sizes $L_x\myeq L_y \myeq 10,\; L_z\myeq 40$ (averaged over
10 independent runs) and  $L_x \myeq L_y \myeq L_z\myeq 40$ (a
sole run). As seen from Fig.~\ref{fig:sigmapeak}, for both system
sizes, results on $\sigmapeak$ are practically identical. Note
that the computation of  $\sigmapeak$ at $\gammadottot \myeq
2.5\times 10^{-6}$ for the large system required about 25 days of
simulation on a 1.8GHz AMD-Athlon CPU. The data point
corresponding to  $\gammadottot \myeq 10^{-6}$ has therefore been
computed using the average over many small systems only. As the
results are not sensitive to the system size, we have used the
smaller system size also in the case of $T\myeq 0.4$ (again
averaging over $10$ independent runs).

For $T\myeq 0.2$, a change in the slope of
$\sigmapeak$-$\gammadottot$-curve is observed at a shear rate of
approximately $\gammadotco\myeq 2.5 \times 10^{-5}$. At shear
rates smaller than $\gammadotco$, the system seems to have enough
time for a partial release of the stress through rearrangements of
particles. Note that the stress overshoot $\sigmapeak$ is observed
at strains smaller than $5\%$. Therefore,  small rearrangements
are sufficient in order to release the stress considerably.
Indeed, an investigation of the mean squared displacement shown in
Fig.~\ref{fig:msd_aging} reveals that the MSD departs from the
plateau for $\tauco \myeq 2 \times 10^{4}$. This time is of the
same order as the inverse of the cross over shear rate thus
suggesting that the cross over in the $\gammadottot$-dependence of
$\sigmapeak$ is related to the beginning of the cage relaxation.
While at higher overall shear rates the response of the system is
dominated by the (shorter) time scale imposed by the shear motion,
it is no longer the case at $\gammadottot<\gammadotco$, where the
inherent system dynamics come into play.
 Although not so
pronounced,  a similar cross over is seen also in the case of
$T\myeq 0.4$ at a larger shear rate in agreement with the
observation that, compared to $T\myeq 0.2$, the MSD at $T\myeq
0.4$ leaves the plateau at a shorter time [see the MSD($T\myeq
0.4$) in Fig.~\ref{msd_y_tw900000_allT}]. Note that, as the
structural relaxation time is approximately proportional to the
age of the system \cite{barratkob}, $\tauco$ is of the order of
$\tw$. The system response  below the crossover is in fact a
complex combination of aging dynamics and stress induced
relaxation. The aging dynamics tends to make the system stiffer
(see below), so that the observed $\sigmapeak$ is higher than the
value one would extrapolate from high shear rates.

The dependence of the stress overshoot  $\sigmapeak$ on the
imposed shear rate is often expressed with a simple formula which
goes back to the Ree-Eyring's viscosity
theory~\cite{Eyring,Larson},
\begin{equation}
\sigmapeak \myeq \sigma_0 + \kB T/v^* \ln(\gammadottot /  \nu_0).
\label{eq:Eyring}
\end{equation}
Here, the \emph{activation volume}, $v^{*}$, is interpreted as the
characteristic volume of a region involved in an elementary shear
motion (hopping) and $\nu_0$ is the attempt frequency of hopping.
Obviously, Eq.~(\ref{eq:Eyring}) makes sense only at high enough
shear rates, for in the case of $\gammadottot < \nu_0$, the second
term on the  right hand side of Eq.~(\ref{eq:Eyring}) becomes
negative. Fitting the data of Fig.~\ref{fig:sigmapeak} to
Eq.~(\ref{eq:Eyring}), we obtain $v^* \myapprox 2.3$ at $T\myeq
0.2$ and $v^* \myapprox 3.0$ at $T\myeq 0.4$. This result is
comparable to the estimates of the free volume from experiments on
polycarbonate, where a value of $v^* \myapprox 3.5 \text{nm}^3$
per segment is reported~\cite{HoHuu}.

Ho Huu and Vu-Khanh~\cite{HoHuu} have extensively studied the
effects of physical aging and strain rate on yielding kinetics of
polycarbonate(PC) for temperatures ranging from $-80^{\circ}{\text
C}$ to $60^{\circ}{\text C}$ [note that $\Tg(\text{PC})\myapprox
140^{\circ}{\text C}$]. In particular, they have measured the
tensile stress at yield point, $\sigmayt$, as a function of strain
rate, $\dot{\epsilon}$, for various temperatures and different
ages of the sample. As for the effect of temperature, they find
that the slope of $\sigmayt(\ln \dot{\epsilon})/T$ (i.e. the
activation volume) is practically independent of $T$. Our data
also show only a weak dependence of $v^*$ on temperature, as
illustrated in the inset of Fig.~\ref{fig:sigmapeak}. Note that we
have also restricted the data-range to higher shear rates where
Eq.~(\ref{eq:Eyring}) is expected to hold better.

The above qualitative agreement on the strain rate dependence of
the stress at yield point for our molecular model glass and
polycarbonate suggests that, for strains smaller than, say $10\%$,
the relevant length scale is that of a segment. In other words,
the chain connectivity has a rather subordinate effect on the
stress at the yield point (in fact, the connectivity becomes
important for larger strains, where the well-known strain
hardening sets in~\cite{Govaert,Johnson}).

For the same binary mixture of Lennard-Jones particles as in the
present work, Rottler and Robbins~\cite{Rottler} studied the
dependence of  $\tauy$, the maximum of the deviatoric stress, on
the shear rate. In contrast to our results, no crossover similar
to that shown in Fig.~\ref{fig:sigmapeak} was observed in this
reference. Furthermore, by varying the temperature in the range of
$T \in [0.01\; 0.3]$ (by a factor of 30), they found that the
slope of the $\tauy$-$\ln \gammadottot$ data did practically not
change with temperature, whereas in our case, as discussed above,
the slope of $\sigmapeak$-$\ln\gammadottot$ approximately scales
with $T$ (see the inset of Fig.~\ref{fig:sigmapeak}). Note,
however, that in Ref.~\cite{Rottler} a smaller cutoff radius of
$\rc\myeq 1.5$ for the Lennard-Jones potential is used, whereas
$\rc \myeq 2.45$ in our model. Furthermore, the pressure
in~\cite{Rottler} is kept at zero at all temperatures, whereas it
is always positive in our simulations. These differences enhance
the repulsive (and therefore athermal) character of the system
simulated by Rottler and Robbins compared to our model. This also
explains why the shear banding is observed at a temperature as low
as $T\myeq 0.01$ in~\cite{Rottler}, whereas we observe it at
$T\myeq 0.2$ and even higher temperatures~\cite{VBBB}. It is also
worth mentioning that the uniaxial strain in  Ref.~\cite{Rottler}
was imposed by a simple instantaneous rescaling of the box
dimension and the positions of all particles, whereas in our case
a more realistic situation is considered: The shear strain in the
fluid is induced through interactions with a moving atomistic
wall. We must however emphasize that, at the present moment, it is
not clear how the above differences in details of the model and in
the applied simulation techniques may lead to the observed
discrepancies in the behavior of the $\sigmapeak$-$\ln
\gammadottot$ curve.

\begin{figure}
\hspace*{-1mm}\epsfig{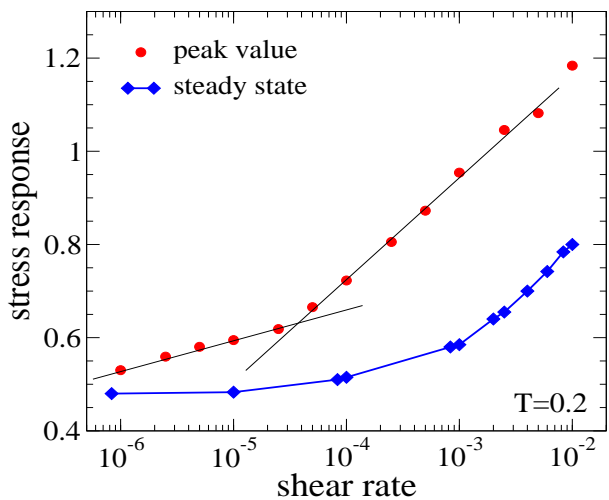}
\caption{\label{fig:sigmapeak+steady} The maximum of the
stress-strain curves as shown in
Fig.~\ref{fig:stress-strain::gammadot_effect}, $\sigmapeak$, and
the steady state stress versus strain rate for a temperature of
$T\myeq 0.2$. For $\sigmapeak$, we average the results for both
system sizes shown in Fig.~\ref{fig:sigmapeak}. }
\end{figure}

As an inspection of Fig.~\ref{fig:stress-strain::gammadot_effect}
reveals, the difference between the peak and the steady state
stresses decreases as $\gammadottot$ is reduced thus suggesting
that, in the limit of vanishing shear rate, $\sigmapeak$ converges
towards the steady state stress (and therefore coincides with the
yield stress that could be extracted from homogeneous flow
experiments). Figure~\ref{fig:sigmapeak+steady} compares these two
quantities, underlining this expectation further.

\begin{figure}
\hspace*{-1mm}\epsfig{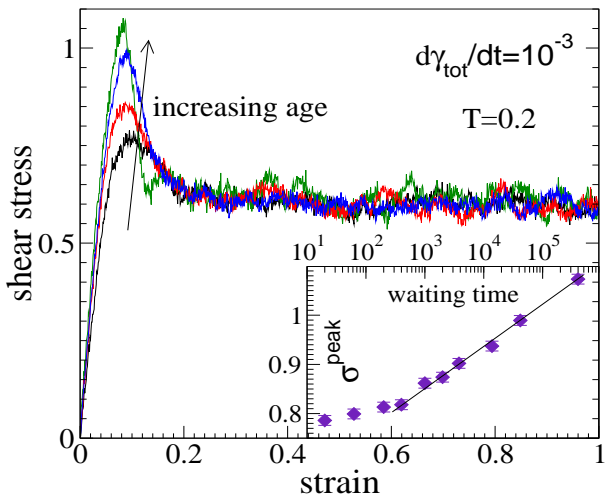}
\caption{\label{fig:stress-strain::tw_effect} Aging effects on the
stress response to an applied strain. $\sigma(\gamma)$ is shown
for a strain rate of $\gammadottot \myeq 10^{-3}$ at $\tw \myeq
20,\; 2\times 10^{3},\; 4.18\times 10^{4}$ and $3.99\times 10^{5}$
LJ time units. At small deformations, the stress increases faster
with progressive aging. The maximum observed stress, $\sigmapeak$,
is reached at smaller strains and is higher for ``older'' samples.
At large deformations, however, the stress converges towards the
same average value regardless of the age of he system. This is a
signature of the recovery of  time translation invariance, or,
equivalently, the erasure of the memory effects due to shear
induced structural relaxation. The inset shows the variation of
the maximum stress with the waiting time. Note that, here, $\tw$
is varied by more than $4$ orders of magnitude, i.e from $\tw\myeq
20$ to $\tw\myeq 3.99\times 10^{5}$. The solid line is a guide for
the eye. The data correspond to a system size of $L_x\myeq
L_y\myeq L_z\myeq 40 $ (a sole run). For the largest waiting time
[$\tw \myeq 3.99\times 10^{5}$ ($\myeq 2\times 10^{7}$ MD steps)],
however, average over $10$ independent runs of a smaller system
size is used ($L_x\myeq L_y\myeq 10$, $L_z\myeq 40$). Note that
already at this (smaller) system size, the size effects are
practically negligible [see also Fig.~\ref{fig:sigmapeak}]. }
\end{figure}

It has been shown in experiments on amorphous polymers like
poly(styrene) and polycarbonate~\cite{Govaert,HoHuu} that aging
strongly alters the response of the system to an applied strain.
At small deformations (below, say $5\%$) the slope of the
stress-strain curve (elastic shear modulus) increases with
progressive aging. Furthermore, the maximum of the stress-strain
curve, $\sigmapeak$, is larger for ``older'' systems and the
subsequent decrease of the  stress (``strain
softening''~\cite{Govaert}) is more pronounced. Similar
observations are also made in experiments on metallic
glasses~\cite{vanAken}. Interestingly,
Fig.~\ref{fig:stress-strain::tw_effect} shows that these features
are not limited to polymers or metallic glasses but can also occur
in simpler models. In Fig.~\ref{fig:stress-strain::tw_effect} the
stress is depicted versus applied strain (defined as $\gamma\myeq
t \gammadottot\myeq t \Uwall/L_z$). Before shearing, the system is
first equilibrated at a temperature of $T\myeq 0.5$. The motion of
the (left) wall is then started at a time $\tw$ after the
temperature quench. Varying $\tw$, we observe similar effects on
the stress response as described above. It is also observed that,
whereas the maximum stress $\sigmapeak$  increases with $\tw$, the
elastic shear modulus (slope of the stress-strain curve) seems to
saturate already for $\tw \geq 2000$ (this is, however, hardly
distinguishable in the scale of the figure).

On the other hand, at large deformations, the stress response does
not show any systematic dependence on the age of the system thus
indicating a recovery of the time translation invariance: steady
shear ``stops aging''~\cite{Kurchan}. In fact, it is well known
that the shear motion promotes structural relaxation and sets an
upper bound ($\sim 1/\gammadottot$) to the corresponding time
scale. Once the steady shear state is reached (which is the case
at deformations comparable to unity), no dependence on the system
age is expected. Results shown in
Fig.~\ref{fig:stress-strain::tw_effect} are also in qualitative
agreement with data reported in Ref.~\cite{Utz}, where the system
response to a homogeneous shear was studied via Monte Carlo
simulations of a binary Lennard-Jones mixture (very close to the
present model). Note that, in Ref.~\cite{Utz}, only the
contribution to the system response of the so called inherent
structure (configurations corresponding to the minima of the
energy landscape) has been considered and the effect of aging is
investigated by applying different cooling rates (not by
``quenching and waiting'' as is the case in our work). Despite
these differences in details, results reported in Ref.~\cite{Utz}
and our observations are quite similar. More quantitative data on
the effect of physical aging on the stress at the yield point is
shown in the inset of  Fig.~\ref{fig:stress-strain::tw_effect}.
Here, $\sigmapeak$ is depicted as a function of the waiting time,
where $\tw$ is varied  by more than four decades. A logarithmic
dependence of $\sigmapeak$ on $\tw$ is clearly seen  for waiting
times larger than a few hundred LJ time units thus covering about
three decades in $\tw$. Such an increase in $\sigmapeak$ is
consistent with the qualitative idea that the system visits
deeper energy minima as aging time increases. A stronger stress is
therefore necessary to overcome the energy barriers towards steady
flow. It is interesting to note that such a $t_w$ dependence of the stress
overshoot is also observed in the SGR model \cite{Sollich}.

As indicated above, simultaneous consideration of figures
\ref{fig:sigmapeak} and \ref{fig:stress-strain::tw_effect}
indicates a rather complex behaviour of $\sigmapeak$ as a function
of $t_w$ and $\gammadottot$. 
Considering the similarity in dependence
for large $\gammadottot$ or large $\tw$, it is tempting to suggest a
rewriting of equation \ref{eq:Eyring} in the 
form $\sigmapeak \myeq \sigma_0 + \kB T/v^* \ln(\gammadottot \tw)$. 
This modified version of Eq.~\ref{eq:Eyring} does, however, not 
describe our data consistently. At $T\myeq 0.2$, for example,
the $\sigmapeak/T$ versus $\ln(\gammadottot \tw)$ curve 
exhibits different slopes for the data obtained by 
varying the imposed shear rate (Fig.~\ref{fig:sigmapeak}) as 
compared to the simulation results where $\tw$ is the adjustable
parameter (corresponding to the data shown in
the inset of Fig.~\ref{fig:stress-strain::tw_effect}).

\begin{figure}
\hspace*{-1mm}\epsfig{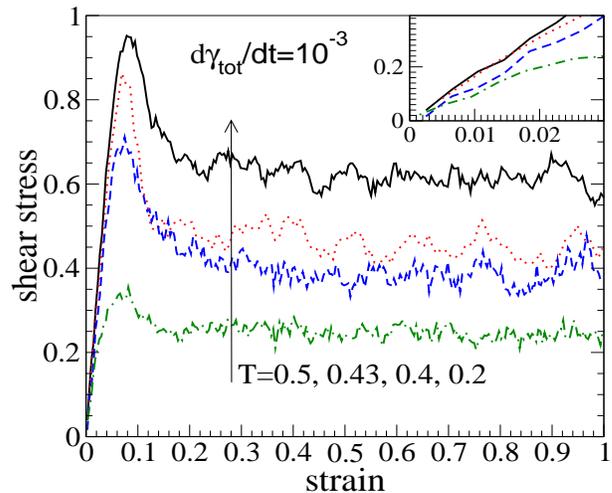}
\caption{\label{fig:stress-strain::T_effect} Effect of temperature
on the stress response to an applied strain. $\sigma(\gamma)$ is
shown for a strain rate of $\gammadottot  \myeq 10^{-3}$ at
$T\myeq 0.2,\; 0.4,\; 0.43$ and $0.5$ (note that the mode coupling
critical temperature of the system is $\Tc\myeq 0.435$). For
temperatures below $\Tc$ the system was first aged during
$\tw\myeq 4\times 10^{4}$ LJ time units before the beginning of
the measurement. Similar to the effect of a decreasing shear rate,
both the maximum and the steady state values of the stress
decrease with increasing temperature. Furthermore, at small
strains, the slope of stress-strain curves (elastic shear modulus)
decreases increasing temperature (see the inset) thus indicating a
softening of the system structure at higher $T$ [compare to
Fig.~\ref{fig:stress-strain::gammadot_effect}]. Note, however,
that these changes are much more pronounced in a narrow
temperature interval around $\Tc$. Results here correspond to
averages over $10$ independent runs. The system size was $L_x\myeq
L_y\myeq 10$ and $L_z\myeq 40$. The inset shows a magnification of
the small strain region of the same data. }
\end{figure}

As for the effect of the temperature on the (transient) stress
response, it is generally known that, due to faster structural
relaxation at higher $T$, the shear stress decreases at higher
temperatures. This is verified in
Fig.~\ref{fig:stress-strain::T_effect} where stress-strain curves
are shown at $T\myeq 0.2,\; 0.4,\; 0.43$ and $0.5$ for a strain
rate of $\gammadottot \myeq 10^{-3}$. Similar to the effect of a
decreasing shear rate, both the maximum and the steady state
values of the stress decrease with increasing temperature.
Furthermore, the slope of stress-strain curves decreases (the
system structure ``softens'') at higher $T$. Qualitatively similar
observations are also made on experimental systems (see, for
example, figure 1.20 in~\cite{Larson}, or
Refs.~\cite{Johnson,Govaert,vanAken,HoHuu}). It is also seen from
Fig.~\ref{fig:stress-strain::T_effect} that a change of
temperature by a factor of two in the glassy state (from $T\myeq
0.2$ to $T\myeq 0.4$) has less impact on the maximum stress,
$\sigmapeak$, than  a smaller $T$-variation close to $\Tc$  (from
$T\myeq 0.4$ to $T\myeq 0.43$). This illustrates the sensitivity
of the yield point to a temperature change in the vicinity of
$\Tc$. Already from this observation, we can expect a similar
impact on the $T$-dependence of the \emph{static} yield stress (see below)
close to the mode coupling critical temperature of the system.
\section {Results at imposed stress}
\label{section:imposedstress}
In this section we study the response of the system to imposed shear stress.
The system is prepared in a similar way as described in previous sections so that,
at the beginning of the measurement, the structural relaxation times
of the system are much larger than the time scale of the simulation.
Starting with $\sigma(t\myeq 0)\myeq 0$, we gradually increase the external stress
(i.e. the force acting  on the atoms of the left wall) and record
quantities of interest, such as the internal energy,
the stress across the system, the center of mass velocity of
the walls and of the fluid, etc...

\begin{figure}
\hspace*{-1mm}\epsfig{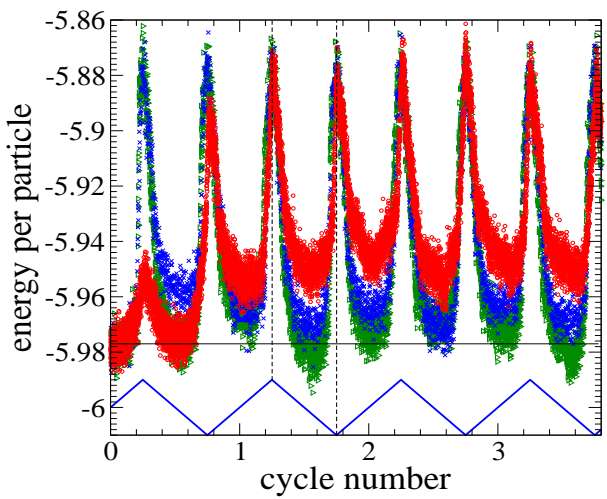}
\caption{\label{fig:epot_versus_t_3rates}
Effect of the rate of stress increase on potential energy per particle. $\epot$
is measured during cyclic variations of the imposed stress as sketched by
the zigzag line (note that $\sigma$  varies in the range $[-0.8\; 0.8]$).
 The horizontal axis counts the number of cycles. Three rates of stress
 variation are shown here, $\sigmadot\myeq 5\times 10^{-4},\; 5\times 10^{-5}$ and $10^{-5}$. T
he higher $\sigmadot$, the higher the potential energy per particle at small
 stresses (minimum of $\epot$). The vertical dashed lines mark simultaneously
the maxima of $\epot$ and $|\sigma|$. They serve to better recognize the asymmetry
of $\epot$ on both sides of the stress maximum and recall the presence of a hysteresis effect.
}
\end{figure}

It is generally accepted that imposing an external
stress leads to a shift in the density of accessible states towards
higher energy configurations. For the binary Lennard-Jones model
of the present work, Fig.~\ref{fig:epot_versus_t_3rates} shows
the potential energy per particle, $\epot$, as measured in
simulations where the imposed stress is periodically varied
the range $\sigma\in [-0.8\; 0.8]$ (see the zigzag line in
 Fig.~\ref{fig:epot_versus_t_3rates}.
Similar stress ramps were also used
by He and Robbins~\cite{He-Robbins} in order to determine
the static friction between two solid bodies mediated by a layer of adsorbed molecules).

Note that the maxima and minima of the potential energy correspond to
$|\sigma|\myeq 0.8$ and $\sigma\myeq 0$ respectively. Starting at a
minimum of $\epot$ ($\sigma\myeq 0$), the potential energy fluctuates
for a while around this minimum before increasing sharply towards a
maximal value. This corresponds to a branch where $|\sigma|$ increases from
$0$ to $0.8$. The descent from this maximum towards the subsequent minimum
($|\sigma|$ decreases from $0.8$ to $0$) is, however, more gradual and
indicates a dependence of $\epot$ on the stress \emph{history}.
Finally, we also observe that, at high $\sigmadot$,
the quiescent energy distribution observed at small stresses at the
very beginning of the stress ramp simulation, is  never reached again
whereas the stress itself passes through zero periodically.
This dependence on $\sigmadot$, however, is considerably
weakened as the stress increase rate reaches values below $5\times 10^{-5}$.

\begin{figure}
\hspace*{-1mm}\epsfig{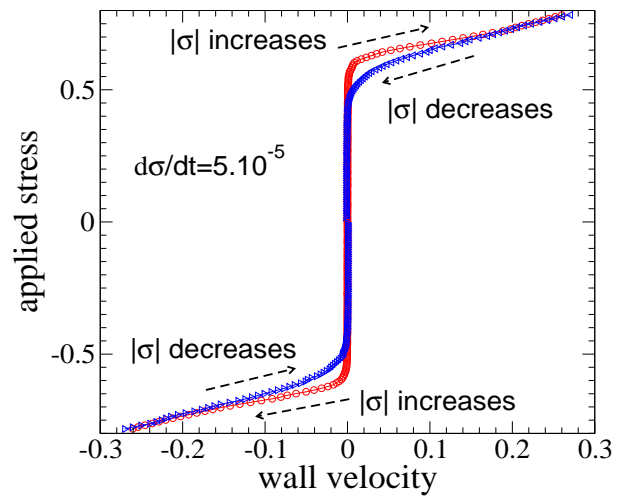}
\caption{\label{fig:Vcm_average_rate5e-5}
The applied shear stress (vertical axis) and the resulting wall velocity
(horizontal axis) measured during stress ramps with a rate of
$\sigmadot \myeq 5\times 10^{-5}$. The result shown here is an
average over two independent runs each containing 15 full cycles of
stress variation [see the zigzag line in Fig.~\ref{fig:epot_versus_t_3rates}].
}
\end{figure}

While the potential energy per particle is easily measured in a simulation,
this is not the case in real experiments. The velocity of the solid boundary
(upon which the stress acts), however, is experimentally accessible.
 Figure~\ref{fig:Vcm_average_rate5e-5} depicts the wall velocity
measured in simulations at $\sigmadot\myeq 5\times 10^{-5}$. Following
the convention, the applied stress is shown on the vertical axis, whereas
on the horizontal axis  the system response is depicted. We first note that,
at small stresses, the system resists to the imposed stress and thus prevents
the wall from moving. Only when the magnitude of the stress exceeds a certain
(yield) value, a non-vanishing wall velocity is observed. Furthermore,
after a cross over regime around the threshold value of the stress, the
wall velocity increases almost linearly with stress increment.

On the other hand,  as the magnitude of the stress is decreased again,
the wall motion first slows down along the same line as in the stress
increase case but then departs towards higher wall velocities. A hysteresis
loop is thus formed as expected from an analysis of the asymmetry of
$\epot$ around the stress maximum [see Fig.~\ref{fig:epot_versus_t_3rates}].
Similar observations are made in experiments on pastes, glass beads,
dense colloidal suspensions~\cite{DaCruz} and foams~\cite{Debregeas::PRL87}.
 Note also that, as expected from the symmetry of the system response with
respect to positive and negative stresses, the shape of the observed
hysteresis loop is identical for both directions (signs) of the applied stress.

\begin{figure}
\hspace*{-1mm}\epsfig{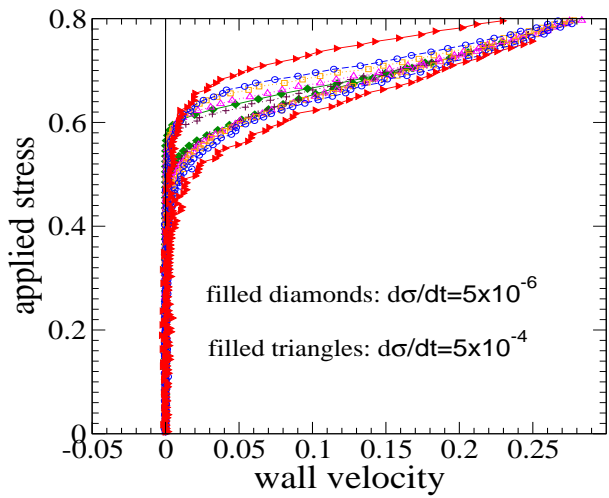}
\caption{\label{fig:Vcm_average_all_rates_inc+dec+5e-6}
The hysteresis loops as measured during stress ramps with rates
of stress variation of $\sigmadot\myeq 5\times 10^{-6}$ (filled diamonds),
$2\times 10^{-5},\; 5\times  10^{-5},\; 10^{-4},\; 2\times 10^{-4}$ and
$5\times 10^{-4}$ (filled triangles). The $\sigmadot
 \myeq 5\times 10^{-6}$-curve is an average over $10$
independent runs with a unique variation of the stress
from $0$ to $0.76$. The remaining curves correspond to
averages over two independent runs each containing many full
cycles of stress variation in the intervall $[-0.8\;\; 0.8]$.
The innermost loop (filled diamonds) corresponds to the smallest
$\sigmadot$. Note that the surface of the hysteresis loop increases
at higher stress variation rates thus indicating stronger retardation
 effects. Note also that, for the two highest $\sigmadot$, the loop
does not close within the simulated stress range. It would close at
much higher stresses than shown in the figure.
}
\end{figure}

Next, we investigate the dependence of the system response to an applied stress
on the rate of stress variation. For this purpose, $\sigmadot$ is varied by
two orders of magnitude, from $5\times 10^{-4}$ to $5\times 10^{-6}$.
Figure~\ref{fig:Vcm_average_all_rates_inc+dec+5e-6} depicts stress
ramp data now averaged using the symmetry with respect to negative
and positive stresses. Again, for all values of $\sigmadot$ shown in
this figure, no flow is observed for too small stresses (below, say $0.4$).
However, for a given stress above, say, $\sigma \myeq 0.7$,  the wall
velocity is \emph{lower} at higher $\sigmadot$. To put it the other way,
when $|\sigma|$ is increased faster, a given wall velocity is reached at
a higher $|\sigma|$, i.e. on a later time. This may be rationalized by noting
that, at a higher stress increase rate, the system has less time to develop a
response corresponding to the actual (instantaneous) stress. Therefore, the
mobility increase corresponding to an increase of the stress is retarded
and is observed later, i.e. at higher stress.

However, it is also seen from Fig.~\ref{fig:Vcm_average_all_rates_inc+dec+5e-6}
that, already at $\sigmadot\leq 2\times 10^{-5}$, the effect of $\sigmadot$ on
the system response is of order of the measurement uncertainty, so
that no systematic dependence on $\sigmadot$ can be seen  for
$\sigmadot\leq 2\times 10^{-5}$. This is consistent with the behavior
of the potential energy per particle which becomes practically independent
of $\sigmadot$ in the same $\sigmadot$-range [see Fig.~\ref{fig:epot_versus_t_3rates}]. Therefore,
we may describe this regime of slow variation of $\sigma$ as quasistatic.

\begin{figure}
\hspace*{-1mm}\epsfig{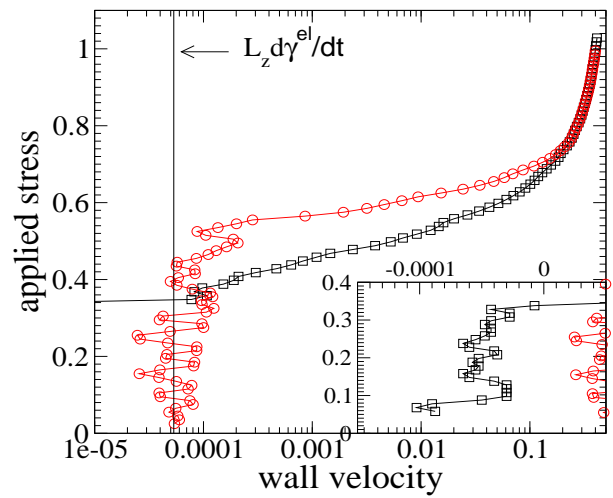}
\caption{\label{fig:Vcm_average_rate2e-5}
The applied shear stress (vertical axis) and the resulting wall velocity
(horizontal axis) measured during stress ramps with a rate of
$\sigmadot \myeq 2\times  10^{-5}$. The vertical solid line roughly
marks the elastic contribution to the strain rate
($\dot{\gamma}^{\text {el}} \myapprox \sigmadot/G$.
Elastic deformation gives rise to a non vanishing wall
velocity even at the smallest imposed stress.
See also Fig.~\ref{fig:stress-strain::gammadot_effect} for
an estimation of the shear modulus $G$). The inset shows that,
as the stress is decreased,  the wall velocity changes its sign
thus indicating that the attractive forces are stronger than the
imposed stress so that the direction of deformation is reversed
in order to reduce the amount of the accumulated strain.
}
\end{figure}

Results presented above and in previous
works~\cite{BB::PRE61::2000,BB2} show that our model
system shares many features of the so called soft glassy
materials. In particular, the existence of a yield stress
is suggested in Figs.~\ref{fig:Vcm_average_rate5e-5}
and~\ref{fig:Vcm_average_all_rates_inc+dec+5e-6}.
Figure~\ref{fig:Vcm_average_rate2e-5} displays further evidence of
the existence of a yield stress still  as the
dramatic change in the wall velocity at a threshold stress value is
emphasized using a logarithmic scale for the horizontal axis.
In a narrow stress range around $\sigma\myeq 0.6$, the wall
velocity and thus the overall shear rate increases approximately
by three orders of magnitude [see also Fig.~\ref{Vcm_Rcm_profile_dz1_sigma0.54-0.70}].
Again, a linear regime is observed at high stresses~\cite{DaCruz}.
Besides the hysteresis already discussed above, an investigation of
the decreasing branch on the stress-wall velocity curve in
Fig.~\ref{fig:Vcm_average_rate2e-5} reveals that, as $\sigma$ falls
below a certain value, the wall velocity becomes even negative [see the inset].
This clearly illustrates the presence of attractive forces which, now,
are stronger than the imposed stress and thus capable of reducing the
amount of strain. Indeed, an inspection of the center of mass position
of the wall and of the fluid shows that both these quantities exhibit
a maximum at the place where the velocity passed through zero
(as the stress is further reduced, $\Xcm$ decreases in accordance
with the observation of a negative velocity). Very similar observations
are also reported on the experimental side~\cite{DaCruz}.

\section{Measurement of the yield stress}
\label{section:yieldstress}
As discussed in section~\ref{section:introduction},
there seems to be a close connection between the existence of a
yield stress and the  observation of the shear banding phenomenon
in many soft glassy materials~\cite{CoussotJRheh46,DaCruz,VBBB}.
In particular, it is commonly expected that, in a state where the
yield stress vanishes (at high temperatures, for example) the shear
bands should also disappear, i.e. the whole system should flow.
In addition to this experimental aspect, a study of the yield
stress is also motivated from the theoretical point of view.
For example, the so called soft glassy rheology model (SGR) of
Sollich~\cite{Sollich2} (an extension of the trap model~\cite{Bouchaud}
taking into account yielding effects due to an external flow) predicts a
linear onset of the dynamic yield stress as the glass transition is approached,
 $\sigma(\gammadottot \to 0) \sim 1-x$. Here, $x$ is a noise temperature,
$x\myeq 1$ corresponds to the glass transition (or ``jamming'') temperature,
and $x<1$ characterizes the glassy or ``jammed'' phase. On the other hand,
numerical studies of a $p$-spin mean field Hamiltonian~\cite{BB::PRE61::2000}
predict that the dynamic yield stress vanishes at all temperatures. There has
recently been a more microscopic approach based on an extension to non
equilibrium situation~\cite{FuchsCates} of the mode coupling theory
of the glass transition (MCT)~\cite{MCTRefs}. An analysis of schematic
models within this approach shows a rather discontinuous change in the
dynamic yield stress at the mode coupling critical temperature, $\Tc$.

The reader may have noticed that the above mentioned theories make predictions
on the \emph{dynamic} yield stress [defined as $\sigma(\gammadottot \to 0)$].
Our interpretation of the shear banding, however, makes use of the idea of
resistance to an applied stress which is related to the presence of a
\emph{static} yield stress. Similar to the difference between the dynamic
and static friction~\cite{RobbinsMueser}, the static and the dynamic yield
stresses are not necessarily identical. Indeed, for our model glass, we find
that $\sigmay>\sigma(\gammadottot \to 0)$ [see Fig.~\ref{fig:fig3}]. Therefore,
a measurement of the static yield stress gives at least an upper bound for the
dynamic counterpart. As we will see below, the static yield stress decreases
rather sharply as the mode coupling critical  temperature of the model
($\Tc \myeq 0.435$) is approached. Unfortunately, when measuring $\sigmay$
at temperatures close to $\Tc$, one is faced with the problem that the time
scale imposed by the external force (which if of order of the inverse stress
variation rate, i.e. $t_{\sigmadot} \myeq \sigma/\sigmadot$) and that of the
(inherent) structural relaxation, $\taurelax$, are not well separated. In particular,
the condition $t_{\sigmadot} \ll \taurelax$ is not valid at temperatures close to $\Tc$.
Therefore, as will be discussed below in more details, a conclusive statement on the
interesting limit of $\sigmay(T\to \Tc)$ can still not be made.

Preliminary results on the static yield stress have been
recently obtained within the driven mean field $p$-spin models~\cite{Berthier}.
Using the fact that the free energy barriers are finite at finite system size,
the model has been investigated by Monte Carlo simulations in the case of $p\myeq 3$ for
a finite number of spins, thus allowing the thermal activations to play a role
which they could not play in the case of an infinite system size. Results of
these simulations support the existence of a critical driving force below which
 the system is trapped ('solid') and above which it is not ('liquid')~\cite{Berthier}.
Results based on this new approach on the temperature dependence of the yield stress and,
in particular, on its behavior close to $\Tc$ are, however, lacking at the moment.

Here, we adopt a method very close to a determination of
 the (static) yield stress in experiments, i.e. we use
the definition of $\sigmay$ as the smallest stress at which
 a flow in the system is observed. As we are interested in a
study of the temperature dependence of $\sigmay$ and, in particular,
in $\sigmay(T)$ close to the mode coupling critical temperature, we
have varied the temperature in the range of $T\in [0.1\; 0.44]$
(recall that $\Tc\myeq 0.435$). For each temperature, $\sigma$
was increased stepwise by an amount of $d \sigma \myeq 0.02$
once in each $1000$ LJ time units during which the velocity
profile corresponding to the imposed stress is measured.
Among other quantities, we also monitor the motion of the
center of mass of the wall and also of the fluid itself. Note
that the overall stress increase rate in these simulations is
$\sigmadot\myeq 2\times 10^{-5}$, and thus corresponds to a
quasi static variation of the stress [see the discussion of
Figs.~\ref{fig:epot_versus_t_3rates} and~\ref{fig:Vcm_average_all_rates_inc+dec+5e-6}].
For each temperature, the simulation was performed using 10
independent initial configurations.

Recall that there is always an elastic contribution to the system response
to an applied stress. The corresponding center of mass velocity can simply
be estimated as $\Vcmel\myeq \sigmadot/G$. This contribution is negligible
at lower $T$ for two reasons: (i) due to the high stiffness of the system
(large $G$), $\Vcmel$ is relatively small and (ii) the onset of the
shear motion is quite sharp at low $T$ thus leading to much higher
velocities (compared to $\Vcmel$) as soon as the applied stress exceeds
$\sigmay$. In contrast, close to $\Tc$,
the shear modulus is quite small [see, for example,
the slope of the stress-strain curve at $T\myeq 0.43$
in Fig.~\ref{fig:stress-strain::T_effect}] thus leading
to a larger $\Vcmel$. Furthermore, there is no sharp variation
in $\Vcm$ as a function of applied stress. For a measurement
of $\sigmay$ close to $\Tc$, it is therefore important to
correct for the elastic contribution to the system response.
For this purpose, we have determined the $T$-dependence of the
shear modulus. The center of mass velocity of the fluid has then
been corrected subtracting, for each temperature, the corresponding
$\Vcmel\myeq \sigmadot/G$.

\begin{figure}
\hspace*{-1mm}\epsfig{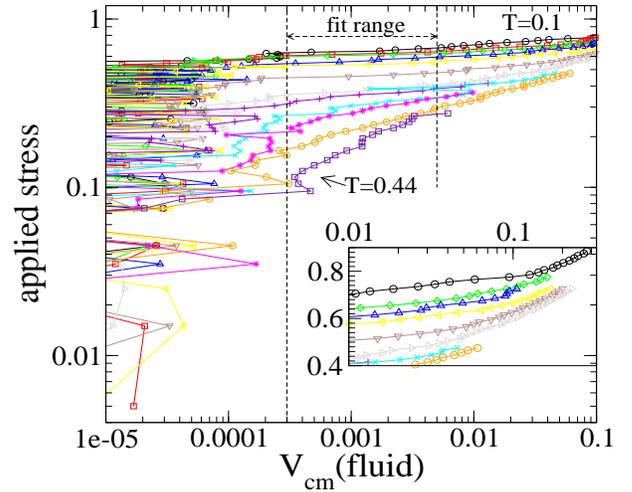}
\caption{\label{fig:Vcm_average_all_rates2e-5_inc_allT}
Effect of the temperature on the response of the system to imposed
shear stress. The imposed stress is shown on the vertical axis
while the center of mass velocity of the fluid (inner part of
the system) is depicted as horizontal axis. Each curve corresponds
to an average over 10 independent runs. The stress was increased
stepwise by an amount of $d\sigma\myeq 0.02$ once in each $dt \myeq 1000$
LJ time units ($\sigmadot\myeq 2\times 10^{-5}$). Note that the
contribution of the elastic deformation to the center of mass
velocity ($\Vcmel \myapprox L_z \sigmadot/(2G)$, using the value
of $G\myapprox 15$ at $T\myeq 0.2$)  has already been subtracted
from the data. Note also that the statistical uncertainty of $\Vcm$
is approximately of order of $10^{-4}$. The vertical dashed lines show
the limits of the $\Vcm$-range used in the fit to $\sigma \myeq \sigmay + a \Vcm$.
The inset is a magnification of the high stress regime.
}
\end{figure}

Figure~\ref{fig:Vcm_average_all_rates2e-5_inc_allT} depicts the
applied stress (vertical axis) and the resulting (corrected)
center of mass velocity of the fluid, $\Vcm$, averaged over
all independent runs (horizontal axis). A log-log plot is used
in order to emphasize the continuous variation of $\Vcm$ with
decreasing stress at high temperatures. Contrary to low temperatures
($T\leq 0.35$) where a plateau followed by a sharp drop towards zero
in $\Vcm$ is observed, the center of mass velocity of the fluid at
high temperatures decreases rather \emph{continuously} for small stresses.

As a first attempt to determine the yield stress, we apply linear fits to
the data shown in Fig.~\ref{fig:Vcm_average_all_rates2e-5_inc_allT}. As
shown in the same figure, the chosen fit range roughly corresponds to
the plateau region at low temperatures. For $T \leq 0.35$, we thus expect
the fit result not to be significantly different from the ``real'' value
of $\sigmay$. However, as an investigation of the high-$T$ behavior of
 $\Vcm$ in Fig.~\ref{fig:Vcm_average_all_rates2e-5_inc_allT} suggests,
this method is not expected to give accurate results for $\sigmay$ at
high temperatures ($T\geq 0.38$, say).

\begin{figure}
\hspace*{-1mm}\epsfig{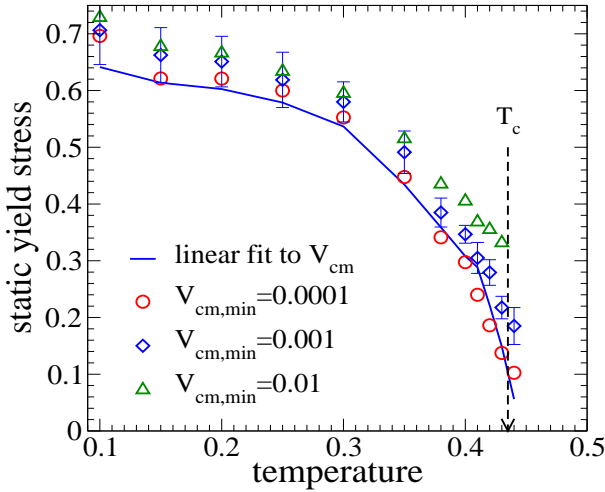}
\caption{\label{fig:yield_stress_versus_T}
The effect of the temperature on the static yield stress, $\sigmay$.
The solid line shows $\sigmay$ obtained from fits to $\sigma \myeq \sigmay+a\Vcm$ ($a$
and $\sigmay$ are fit parameters) using the fit range $\Vcm\in [0.0003\; 0.005]$
[see Fig.~\ref{fig:Vcm_average_all_rates2e-5_inc_allT}]. The symbols
correspond to $\sigmay$ defined as the smallest stress for which
(and for all subsequent higher stresses) the wall velocity exceeds
a certain minimum value, $\Vcmmin$. Three choices of $\Vcmmin$ are
compared: $10^{-4}$ (circles),  $10^{-3}$ (diamonds) and  $10^{-2}$ (triangles).
While $\sigmay$ is relatively insensitive to a choice of $\Vcmmin$ at
low temperatures, it is not the case for temperatures close to $\Tc$,
where it continuously decreases as the $\Vcmmin$ is reduced. The vertical
arrow marks the mode coupling critical temperature $\Tc\myeq 0.435$.
For clarity, error bars are shown for the case of $\Vcmmin\myeq 10^{-3}$ only.
}
\end{figure}

A slightly different approach in determining $\sigmay$ is to find
the smallest stress for which the center of mass velocity exceeds a
certain, small value, $\Vcmmin$. Here, we further require that
$\Vcm$ must \emph{remain} larger than $\Vcmmin$ for all subsequent
stresses. This last condition serves to reduce errors due to
fluctuations of $\Vcm$. In applying this definition, we use the
result of each independent run on $\Vcm$ separately and thus obtain,
for each $\Vcmmin$, a set of yield stress values. This allows an
estimate of the statistical error.  Figure~\ref{fig:yield_stress_versus_T}
compares the yield stress obtained via the linear fit to $\Vcm$ with results
of the second approach for $\Vcmmin\myeq 10^{-4},\; 10^{-3}$ and $10^{-2}$.
Not unexpectedly, it is seen from  Fig.~\ref{fig:yield_stress_versus_T} that
the quality of results on $\sigmay$ strongly depends on temperature. At
temperature far enough from $\Tc$, say, for $T<0.35$, $\sigmay$ is rather 
insensitive to a change of $\Vcmmin$ (differences caused
by various choices of $\Vcmmin$ are of order of the statistical error).
At higher temperatures, however, the variation of $\sigmay$ with the
choice of $\Vcmmin$ is remarquable: At $T\myeq 0.43$ it varies between
$0.14$, $0.22$ and $0.33$ for $\Vcmmin\myeq 10^{-4},\; 10^{-3}$ and $10^{-2}$.
Therefore, result on $\sigmay$ at temperatures close
to $\Tc$ should be considered as rough estimates only.

\begin{figure}
\hspace*{-1mm}\epsfig{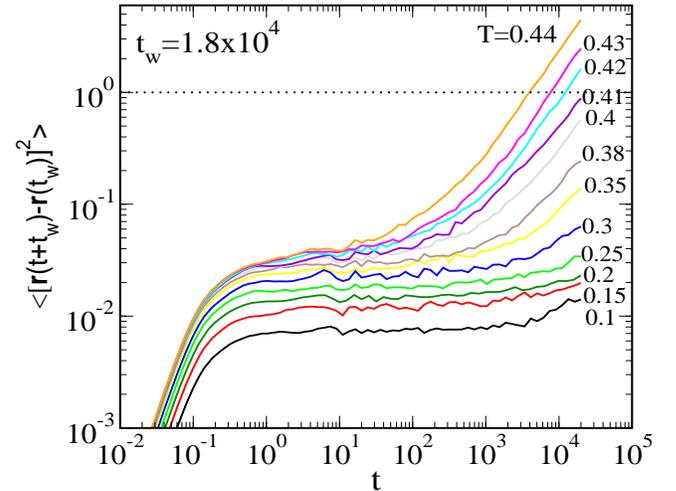}
\caption{\label{msd_y_tw900000_allT}
The mean square displacement (MSD) of a tagged particle averaged over
 all three spatial directions at various temperatures ranging from the
supercooled state ($T\myeq 0.44 > \Tc\myeq 0.435$) down to the frozen
state $T\myeq 0.1$. The time $\tw$ indicates the time elapsed between
the temperature quench and the beginning of the measurement. At a time
of $t\myeq 2 \times 10^{4}$, the MSD hardly leaves the plateau at low
temperatures. For temperatures $T \geq 0.41$, however,  it approximately
reaches the size of a particle within the same time interval indicating
that a complete structural relaxation has taken place. The horizontal
dotted line marks $\text{MSD}\myeq 1$.
}
\end{figure}

The origin of the difficulty in estimating the static yield stress of
the system at temperatures close to $\Tc$, can be understood by comparing
the time scales relevant to the problem. First, there is a time scale
related to the imposed stress $t_{\sigmadot} \myeq \sigma / \sigmadot$.
The second relevant  time scale is that of the structural relaxation, $\taurelax$.
The static yield stress is well defined in the limit of a quasi static variation
of stress, i.e. $\sigmadot\to 0$ ($t_{\sigmadot}\to \infty$) while at the
same time keeping $\taurelax\gg t_{\sigmadot}$. Using $\sigma\myapprox 0.5$
and $\sigmadot \myeq 2 \times 10^{-5}$ (note that this value of $\sigmadot$
was used at all temperatures in order to determine the yield stress) we
obtain $t_{\sigmadot} \myapprox 2 \times 10^{4}$. We are therefore led
to verify if the condition $\taurelax\gg 2 \times 10^4$ is satisfied at
all temperatures. For this purpose, we define $\taurelax$ as the time
needed by the mean square displacement of a tagged particle to reach
the particle size. Figure~\ref{msd_y_tw900000_allT} shows the mean
square displacement of the \emph{unsheared} system for $T\in [0.1\; 0.44]$
(recall that $\Tc\myeq 0.435$). For all these temperatures, the waiting
time between the temperature quench (from an initial temperature of
$T\myeq 0.5$ to the actual temperature) and the beginning of the
measurement was $\tw \myeq 1.8 \times 10^{4}$. At low temperatures,
the MSD practically remains on a plateau for the whole duration of
the simulation indicating that $\taurelax$ is much larger than the
simulated time of $2\times 10^4$ LJ time units. At higher temperatures
($T\geq 0.41$), however,  after a long plateau, the
MSD eventually enters the diffusive regime and reaches a value comparable
to unity within the simulated time window. Obviously the condition
$\taurelax \gg t_{\sigmadot}$ is violated at these temperatures.
Hence at least for a waiting time of $\tw\myeq 4\times 10^4$ and
for the choice of $\sigmadot\myeq 2\times 10^{-5}$, the computed
static yield stress is not well defined close to $\Tc$.

\section{Conclusion}
\label{section:conclusion}
Results on the yield behavior of a model glass (a 80:20 binary Lennard-Jones
mixtures~\cite{Kob-Andersen}), studied by means of molecular
dynamics simulations, have been  reported. One of the major motivations
of the present work is the observation of shear localization
(below the glass transition temperature and at low shear rates)
in the present model and the suggestion of a link between this
phenomenon and the existence of a static yield stress~\cite{VBBB}
(under applied stress, the system does not flow until the stress
exceeds a threshold value). A particular emphasis thus lies on
the yield stress and its dependence on temperature.

First, the system stress-strain curve  under startup of steady shear has been studied.
The effect of physical aging (characterized by the waiting
time, $\tw$), shear rate ($\gammadottot$), and temperature
on the stress-strain relation has been  investigated. Regardless
of these parameteres, all observed stress-strain curves
first exhibit an elastic regime at small shear deformations ($\gamma \leq 0.02$).
The stress then increases up to a maximum, $\sigmapeak$, before decreasing
towards the steady state stress at large deformations.
 The steady state stress
(corresponding to large deformations) shows a dependence on temperature
and on the applied shear rate, but is independent of the system history,
indicating a recovery of the time translation invariance  due to shear
induced structural relaxation~\cite{BB::PRE61::2000}. In contrast, the stress overshoot
 $\sigmapeak$,
(the first maximum of the stress-strain curves, spmetimes described as a dynamical yield
stress) depends on the imposed shear rate
and on the waiting time (physical aging). It is observed that,
at relatively high shear rates or for large waiting times, the
maximum stress increases with $\ln(\gammadottot)$ or with  $\ln(\tw)$,
respectively. These observations are consistent with  experiments on
amorphous polymers~\cite{Govaert,vanAken} and on metallic glasses~\cite{HoHuu,Johnson},
and also correspond to the behaviour predicted using the soft glassy rheology model \cite{Sollich}.

For shear rates below a certain, cross over shear rate, $\gammadotco$,
however, a decrease in the slope of $\sigmapeak$-$\gammadottot$ curve
is seen. A comparison with the steady state shear stress  suggests that
$\sigmapeak$ saturates at the steady state stress level as the imposed
shear rate approaches zero. Moreover, an analysis of the mean square
displacements of the unsheared system reveals that the  cross over
shear rate, $\gammadotco$, is very close to $1/\tauco$, where
$\tauco$ marks the time  for which the mean square displacement
gradually departs from the plateau-regime [see Fig.~\ref{fig:msd_aging}].
We therefore associate  this crossover with the beginning of the cage relaxation,
which leads to the possibility of small (compared to the size of a particle)
rearrangements thus allowing at least a partial release of the stress.
$gammadotco$ is also comparable to the inverse of the waiting time:
for $gammadottot < gammadotco$, the response of the system is directly influenced by the aging dynamics.

In order to build a closer connection between our studies and
typical rheological experiments, stress ramp simulations are
performed and the system response is analyzed for stress
increase rates ranging from $\sigmadot \myeq 5\times 10^{-4}$ to
 $\sigmadot \myeq 5\times 10^{-6}$. In agreement with experiments
on complex systems like pastes, dense colloidal suspensions~\cite{DaCruz}
and foams~\cite{Debregeas::PRL87}, hysteresis loops in the system response
are observed. These loops become wider as $\sigmadot$ increases. An
analysis of the potential energy per particle for different $\sigmadot$
nicely shows how  high energy configurations are favored by the faster
stress variations. This also yields an estimate of quasi static stress application.
 We find that, for our model, $\sigmadot \myeq 2\times 10^{-5}$ is slow
enough so that simulations with this stress variation rate can be used
in order to obtain a reliable estimate of the static yield stress.

Finally, the static yield stress, $\sigmay$, is determined and its reliability
is discussed. Our numerical results confirm the observation of   reference \cite{VBBB},
that the static yield stress is higher than the low shear rate limit
$\sigma(\gammadot\rightarrow 0)$ observed in steady shear experiments. The system can therefore
produce shear bands for stresses in the range [$\sigma(\gammadot\rightarrow 0)$,$\sigmay$].

At temperatures far below the mode coupling critical temperature
of the model ($\Tc\myeq 0.435$), a slight increase of $\sigmay$ with further
cooling is observed.
At temperatures close to $\Tc$, however, the static yield
stress strongly decreases as $T$ is increased towards $\Tc$. As to the
reliability of the data, relatively accurate estimate of $\sigmay$ is
obtained at low temperatures (for $T\leq 0.35$). Results on the yield stress
at temperatures close to $\Tc$, however, are very sensitive to the applied
criterion. An investigation of the dynamics of the unperturbed system reveals
that, for $T$ close to $\Tc$, the structural relaxation times are far from being
large compared to the time scale imposed by the external force (the inverse of the
stress increase rate, $ \sigma / \sigmadot$). Therefore, for the simulated waiting
time of $4\times 10^4$, the static yield stress is no longer well defined at these
high temperatures. This underlines the fact that a very good separation of time scales
between the experimental and intrinsic time scales is necessary in order to properly define
a static yield stress.

It must, however, be emphasized that, even though an increase of $\sigmadot$
apparently leads to a validity of  $\taurelax \gg t_{\sigmadot}$, this would
violate the condition of a quasi static variation of the stress. A more physical
way to improve the accuracy of results on $\sigmay$ is to increase the waiting time,
in order to allow $\taurelax$ to grow beyond $t_{\sigmadot}$. Noting that, at higher
temperatures (but still below $\Tc$), $\taurelax$ increases less strongly with $\tw$
(interrupted aging),
the limit of large $\taurelax$ becomes progressively more time consuming in terms of
computation time.

Our numerical study shows that a very simple model,
studied numerically on relatively short time scales,
can exhibit most of the complex rheological behaviour of
soft glassy systems, but also of ''hard'' (metallic)
glasses (it is interesting in this respect to note that
the simulated system was originally intended to mimic
a NiPd metallic glass). This suggests that these features are
generic to most glassy systems, although in practice the values of the
parameters may considerably vary from system to system.

\section*{Acknowledgments}
We thank L. Berthier, M. Fuchs and A. Tanguy for useful
discussions. F.V. is supported by the Deutsche
Forschungsgemeinschaft (DFG), Grant No VA 205/1-1. Generous grants
of simulation time by the ZDV-Mainz and PSMN-Lyon and IDRIS
(project No 031668-CP: 9) are also acknowledged.


\begin{references}

\bibitem{Kob-Andersen}
W. Kob and H.C. Andersen, Phys. Rev. E {\bf 52}, 4134 (1995);
W.~Kob, J.~Phys.:~Condens.~Matter {\bf 11}, R85 (1999), and
references therein. Our unit of time is
$\sqrt{m\sigma^2/\epsilon}$, differing by a factor $\sqrt{48}$
from the one used in this reference. Note also that confinement
effects can modify the time scale for structural relaxation, and
therefore the glass transition temperature as discussed by P.
Scheidler et al., Europhys. Lett., in press (2002) for the present
model and by F. Varnik et al., Phys. Rev. E {\bf 65}, 021507
(2002) for a confined polymer melt. The temperatures we consider
are low enough that this effect can be ignored for layers not too
close to the walls (typically at distances $z-z_{\text{wall}} \ge
2$ from the wall).

\bibitem{VBBB} F. Varnik, L. Bocquet, J.-L. Barrat, L. Berthier
Phys. Rev. Lett. {\bf 90}, 095702 (2003).

\bibitem{Govaert}
E.M. Arruda, M.C. Boyce and R. Jayachandran, Mechanics of Materials {\bf 19}, 193 (1995);
L.E. Govaert, H.G.H. van Melick and H.E.H. Meijer, polymer {\bf 42}, 1271 (2001);
H.G.H. van Melick, L.E. Govaert and H.E.H. Meijer, polymer {\bf 44}, 3579 (2003).

\bibitem{vanAken} B. van Aken, P. de Hey and J. Sietsma,
Materials Science \& Engineering A {\bf 278}, 247 (2000).

\bibitem{HoHuu} C. Ho Huu and T. Vu-Khanh,
Theoretical and Applied Fracture Mechanics {\bf 40}, 75 (2003).

\bibitem{Johnson} W.L. Johnson, J. Lu and M.D. Demetriou,
Intermetallics {\bf 10}, 1039 (2002).

\bibitem{DaCruz} F. Da Cruz, F. Chevoir, D. Bonn, P. Coussot
Phys. Rev. E {\bf 66} 051305 (2002).

\bibitem{Debregeas::PRL87}
G. Debr\'egeas, H. Tabuteau, and J.-M. di Meglio,
Phys. Rev. Lett. {\bf 87}, 178305 (2001).

\bibitem{Sollich} P. Sollich, F. Lequeux, P. H\'ebraud, M.E. Cates,
Phys. Rev. Lett. {\bf 78}, 2020 (1997);
S. M. Fielding, P. Sollich, M. E. Cates, J. Rheol., {\bf 44}, 323 (2000).

\bibitem{Bonn2}
D. Bonn, P. Coussot, H.T. Huynh and F. Bertrand, Europhys. Lett. {\bf 59}, 786 (2002).

\bibitem{Coussot-Raynaud-et-al::PRL88::2002}
P. Coussot {\it et al}, Phys. Rev. Lett. {\bf 88}, 218301 (2002).

\bibitem{Larson} R.G. Larson,
\emph{The structure and Rheology of Complex Fluids} (Oxford University Press, New York) 1999.

\bibitem{Bonn}
D. Bonn, S. Tanase, B. Abou, H. Tanaka and J. Meunier, Phys. Rev. Lett. {\bf 89}, 015701 (2002).

\bibitem{Pignon::JRheo40::1996}
F. Pignon, A. Magnin, and J.-M. Piau, J. Rheol., {\bf 40}, 573 (1996).


\bibitem{barratkob} W. Kob and J.-L. Barrat,
Phys. Rev. Lett. {\bf 78}, 4581 (1997); Europhys. Lett. {\bf 46},
637 (1999); Eur. Phys. J. B {\bf 13}, 319 (2000).
\bibitem{BB::PRE61::2000}
L. Berthier, J.-L. Barrat and J. Kurchan, Phys. Rev. E {\bf 61}, 5464 (2000);
J.-L. Barrat and L. Berthier, Phys. Rev. E {\bf 63}, 012503 (2000).

\bibitem{Kurchan}
J. Kurchan cond-mat/9812347 (1998), cond-mat/0011110 (2000);

\bibitem{BB2}
L. Berthier and J.-L. Barrat, J. Chem. Phys. {\bf 116}, 6228 (2002);
Phys. Rev. Lett. {\bf 89}, 095702 (2002).

\bibitem{Lacks} See, however,
D.J. Lacks, Phys. Rev. Lett. {\bf 87}, 225502 (2001),
for an investigation of shear thinning in terms of potential energy landscape.

\bibitem{olmsted}
See e.g. C.-Y.David~Lu, P.D. Olmsted, R.C. Ball, Phys. Rev. Lett. {\bf 84}, 642 (2000)
which also contains many references to relevant experimental work.

\bibitem{Dhont} J.K.G. Dhont, Phys. Rev. E {\bf 60}, 4534 (1999);
X.F. Yuan, Europhys. Lett. {\bf 46}, 542 (1999).

\bibitem{Chen}
L.J. Chen, B.J. Ackerson, C.F. Zulowski, J. Rheol, {\bf 38}, 193 (1993).

\bibitem{Losert::PRL85::2000}
W. Losert, L. Bocquet, T.C. Lubensky and J.P. Gollub,
Phys. Rev. Lett. {\bf 85}, 1428 (2000).


\bibitem{footnote1} If there is no such region,
the yield stress is necessarily \emph{smaller}
than the steady state stress at the imposed shear rate.
This, however, is impossible, for the startup of the shear motion in
response to an imposed stress leads to at least a partial release of the stress.

\bibitem{CoussotJRheh46}
P. Coussot, Q.D. Nguyen H.T. Huynh and D. Bonn, J. Rheol. {\bf 46}, 573 (2002).

\bibitem{Lemaitre} A. Lemaitre, preprint cond-mat/0206417.

\bibitem{Derec} C. D\'erec, A. Ajdari, G. Ducouret and F. Lequeux, C.R. Acad. Sci. Paris IV {\bf 1} 1115 (2000);
C. D\'erec, A. Ajdari and F. Lequeux Eur. Phys. J. E {\bf 4} 355 (2001).

\bibitem{Picard} G. Picard, A. Ajdari, L. Bocquet, F. Lequeux, Phys. Rev. E {\bf 66}, 051501 ( 2002).


\bibitem{He-Robbins}
G. He and M.O. Robbins Phys. Rev. B. {\bf 64} 035413 (2001).

\bibitem{footnote2}  Note that, by choosing a smaller integration time step and applying the rescaling procedure more frequently, one can increase the effective thermostating rate. However, a thermostating rate larger than $1/\tau_{\text{VACF}}$, ($\tau_{\text{VACF}}$ is the decay time of the velocity auto correlation function) does not make much sense as the heat transferred to the $x$-component of the particle velocities must be first dissipated to the other directions before it can be removed by a rescaling of, say $y$-component of the particle velocities.

\bibitem{Utz} M. Utz, P.G. Debenedetti and F.H. Stillinger, Phys. Rev. Lett. {\bf 84} 1471.

\bibitem{Khan} H.J. Walls, S.B. Caines, A.M. Sanchez and  S.A.
Khan, Journal of Rheology, {\bf 47}, 847 (203) and references
therein.

\bibitem{Rottler} J\"org Rottler and Mark O. Robbins, cond-mat/0303276.

\bibitem{Eyring} H. Eyring, J. Chem. Phys. {\bf 4}, 283 (1936);
T. Ree and H. Eyring in: F.R. Eirich(Ed.),
Rheology, vol. II, (Academic Press, New York, 1958), pp. 83-144 (Chapter III).



\bibitem{Sollich2} P. Sollich, Phys. Rev. E {\bf 58}, 738 (1998).

\bibitem{Bouchaud} J.P. Bouchaud, J. Phys. I {\bf 2}, 1705 (1992);
C. Monthus and J.P. Bouchaud, J. Phys. A {\bf 29}, 3847 (1996).


\bibitem{FuchsCates} M. Fuchs and M.E. Cates, Phys. Rev. Lett. {\bf 89}, 248304 (2002);
 M. Fuchs and M.E. Cates, Faraday Discuss. {\bf 123}, 267-286 (2003) [see also cond-mat/0210321 and cond-mat/0207530].

\bibitem{MCTRefs}
U. Bengtzelius, W. G\"otze, and A. Sj\"olander, J. Phys. C {\bf 17},  59115  (1984);
E. Leutheusser, Phys. Rev. A, {\bf 29}, 2765 (1984).
W. G\"otze,  in {\em Les Houches 1989, Session LI}, edited by J.~P. Hansen,
D. Levesque, and J. Zinn-Justin (North-Holland, Amsterdam, 1989).


\bibitem{RobbinsMueser} M. O. Robbins and M. H. M\"user, in Modern Tribology Handbook,
Edited by B. Bhushan (CRC Press, Boca Raton, 2001) (cond-mat/0001056).

\bibitem{Berthier} L. Berthier, J. Phys. C 15, S933 (2003).
\end{references}
\end{document}